\documentclass[12pt]{article}

\usepackage{amsfonts}
\usepackage[centertags]{amsmath}
\usepackage{amssymb}
\usepackage[verbose,colorlinks=true,naturalnames=true,linkcolor=blue,]{hyperref}
\usepackage{ulem}
\usepackage{xcolor}

\def\ba{\begin{array}}
\def\ea{\end{array}}
\def\be{\begin{equation}}
\def\ee{\end{equation}}
\def\bea{\begin{eqnarray}}
\def\eea{\end{eqnarray}}
\usepackage[verbose,colorlinks=true,naturalnames=true,linkcolor=blue,]{hyperref}

\newcounter{rown}

\begin{document}

\title{Quantizations of $D=3$ Lorentz symmetry 
} 
\author{J. Lukierski$^{1}$ and V.N. Tolstoy$^{1,2}$ \\
\\
$^{1}$Institute for Theoretical Physics,  \\
University of Wroc{\l}aw, pl. Maxa Borna 9, \\
50--205 Wroc{\l }aw, Poland\\
\\
$^{2}$Lomonosov Moscow State University,\\
Skobeltsyn Institute of Nuclear Physics, \\
Moscow 119991, Russian Federation}
\date{}
\maketitle

\begin{abstract}
Using the isomorphism $\mathfrak{o}(3;\mathbb{C})\simeq\mathfrak{sl}(2;\mathbb{C})$ we develop a new simple algebraic technique for complete classification of quantum deformations (the classical $r$-matrices) for real forms  $\mathfrak{o}(3)$ and $\mathfrak{o}(2,1)$ of the complex Lie algebra $\mathfrak{o}(3;\mathbb{C})$ in terms of real forms of $\mathfrak{sl}(2;\mathbb{C})$: $\mathfrak{su}(2)$, $\mathfrak{su}(1,1)$ and $\mathfrak{sl}(2;\mathbb{R})$.  We prove that the  $D=3$ Lorentz symmetry $\mathfrak{o}(2,1)\simeq\mathfrak{su}(1,1)\simeq\mathfrak{sl}(2;\mathbb{R})$ has three different Hopf-algebraic quantum deformations which are expressed in the simplest way by two standard $\mathfrak{su}(1,1)$ and $\mathfrak{sl}(2;\mathbb{R})$ $q$-analogs and by simple Jordanian $\mathfrak{sl}(2;\mathbb{R})$ twist deformations. These quantizations are presented in terms of the quantum Cartan-Weyl generators for the quantized algebras $\mathfrak{su}(1,1)$ and $\mathfrak{sl}(2;\mathbb{R})$ as well as in terms of quantum Cartesian generators for the quantized algebra $\mathfrak{o}(2,1)$. Finaly, some applications of the deformed $D=3$ Lorentz symmetry are mentioned.
\end{abstract}

\setcounter{equation}{0} 
\section{Introduction}
The search for quantum gravity is linked with studies of noncommutative space-times and quantum deformations of space-time symmetries. The considerations of simple dynamical models in quantized gravitational background (see e.g. \cite{FrLi2006,CiKoPrRo2016}) indicate that the presence of quantum gravity effects generates noncommutativity of space-time coordinates, and as well the Lie-algebraic space-time symmetries (e.g. Lorentz, Poincar\'{e}) are modified into quantum symmetries, described by noncocommutative Hopf algebras, named after Drinfeld quantum deformations or quantum group \cite{Dr1986}. We recall that in relativistic theories the basic role is played by Lorentz symmetries and Lorentz algebra, i.e. all aspects of their quantum deformations should be studied  in very detailed and careful way. 

For classifications, constructions and applications of quantum Hopf deformations of an universal enveloping algebra $U(\mathfrak{g})$ of a Lie algebra $\mathfrak{g}$, Lie bialgebras ($\mathfrak{g},\delta$) play an essential role (see e.g. \cite{Dr1986, EtKa1996} and \cite{ChPr1994, Ma1995}). Here the {\it cobracket} $\delta$ is a linear skew-symmetric map $\mathfrak{g}\rightarrow\mathfrak{g}\wedge\mathfrak{g}$ with the relations consisted with the Lie bracket in $\mathfrak{g}$:
\begin{eqnarray}
\begin{array}{rcl}\label{in1}
&&\delta([x,y])\;=\;[x\otimes1+1\otimes x,\delta(y)]-[y\otimes1+1\otimes y,\delta(x)],
\\[7pt]
&&(\delta\otimes\mathop{\rm id})\delta(x)+{\rm cycle}\;=\;0
\end{array}
\end{eqnarray}
for any $x,y\in\mathfrak{g}$. The first relation in (\ref{in1}) is a condition of the 1-cocycle and the second one is the co-Jacobi identity (see \cite{Dr1986, Ma1995}). The Lie bialgebra ($\mathfrak{g},\delta$) is a correct infinitesimalisation of the quantum Hopf deformation  of $U(\mathfrak{g})$ and the operation $\delta$ is a an infinitesimal part of difference between a coproduct $\Delta$ and an oposite coproduct $\tilde{\Delta}$ in the Hopf algebra, $\delta(x)=h^{-1}(\Delta-\tilde{\Delta})\mod h$ where $h$ is a deformation parameter. Any two Lie bialgebras ($\mathfrak{g},\delta$) and ($\mathfrak{g},\delta'$) are isomorphic (equivalent) if they are connected by a $\mathfrak{g}$-automorphism $\varphi$ satisfying the condition 
\begin{eqnarray}\label{in2}
\delta(x)=(\varphi\otimes\varphi)\delta'(\varphi^{-1}(x)) 
\end{eqnarray}
for any $x\in\mathfrak{g}$. Of special our interest here are the quasitriangle Lie bialgebras ($\mathfrak{g},\delta_{(r)}$):=($\mathfrak{g},\delta,r$), where the cobracket $\delta_{(r)}$ is given by the classical $r$-matrix $r\in\mathfrak{g}\wedge\mathfrak{g}$ as follows:
\begin{eqnarray}\label{in3}
\delta_{(r)}(x)\!\!&=\!\!&[x\otimes1+1\otimes x,r]. 
\end{eqnarray}
It is easy to see from (\ref{in2}) and (\ref{in3}) that \textit{two quasitriangular Lie bialgebras ($\mathfrak{g},\delta_{(r)}$) and ($\mathfrak{g}, \delta_{(r')}$) are isomorphic iff the classical $r$-matrices $r$ and $r'$ are isomorphic, i.e. $(\varphi\otimes\varphi)r'=r$}. Therefore for a classification of all nonequivalent quasitriangular Lie bialgebras ($\mathfrak{g},\delta_{(r)}$) of the given Lie algebra $\mathfrak{g}$ we need find all nonequivalent (nonisomorphic) classical $r$-matrices. Because nonequivalent quasitriangular Lie bialgebras uniquely determine non-equivalent quasitriangular quantum deformations (Hopf algebras) of $U(\mathfrak{g})$ (see \cite{Dr1986, EtKa1996}) therefore the classification of all nonequivalent quasitriangular Hopf algebras is reduced to the classification of the nonequivalent classical $r$-matrices.
 
In this paper we investigate the quantum deformations of $D=3$ Lorentz symmetry. Firstly, following \cite{BoLuTo2016}, we obtain the complete classifications of the nonequivalent (nonisomorphic) classical $r$-matrices for complex Lie algebra $\mathfrak{sl}(2;\mathbb{C})$ and its real forms $\mathfrak{su}(2)$, $\mathfrak{su}(1,1)$ and $\mathfrak{sl}(2;\mathbb{R})$ with the help of explicite formulas for the automorphisms of these Lie algebras in terms of the Cartan-Weyl bases. In the case of $\mathfrak{sl}(2;\mathbb{C})$ there are two noniquivalent classical $r$-matrices - standard and Jordanian ones. For $\mathfrak{su}(2)$ algebra there is only the standard nonequivalent $r$-matrix. These results are well known. For the $\mathfrak{su}(1,1)$ case we obtained three noneqvivalent $r$-matrices - standard, quasi-standard and quasi-Jordanian ones. In the case of $\mathfrak{sl}(2;\mathbb{R})$ we find also three noneqvivalent $r$-matrices - standard, quasi-standard and Jordanian ones. Then using isomorphisms $\mathfrak{o}(2,1)\simeq\mathfrak{su}(1,1)\simeq\mathfrak{sl}(2;\mathbb{R})$ we express these $r$-matrices in terms of the Cartesian basis of the $D=3$ Lorentz algebra $\mathfrak{o}(2,1)$ and we see that two systems with three $r$-matrices for $\mathfrak{su}(1,1)$ and $\mathfrak{sl}(2;\mathbb{R})$ algebras coincides. Thus we obtain that the isomorphic Lie algebras $\mathfrak{su}(1,1)$ and $\mathfrak{sl}(2;\mathbb{R})$ have the isomorphic systems of their quasitriangle Lie bealgebras. In the case of $\mathfrak{o}(2,1)$ we obtain that the $D=3$ Lorentz algebra has two standard $q$-deformations and one Jordanian. These Hopf deformations are presented in explicite form in terms of the quantum Cartan-Weyl generators for the quantized universal enveloping algebras of $\mathfrak{su}(1,1)$ and $\mathfrak{sl}(2;\mathbb{R})$ and also in the terms of the quantum Cartesian generators. 

It should be noted that the full list of the noniquivalent classical $r$-matrices for $\mathfrak{sl}(2;\mathbb{R})$ and $\mathfrak{o}(2,1)$ Lie algebras has been obtained early by different methods \cite{Re1996, Go2000} 
(see also \cite{ReHeRa2005, BaBlMu2012, BaMeNa2017}), however the complete list of the nonequivalent Hopf quantisations for these Lie algebras has not been presented in the literature. Furthermore, there was put forward an incorrect hypothesis that the isomorphic Lie algebra $\mathfrak{su}(1,1)$ and $\mathfrak{sl}(2;\mathbb{R})$ do not  have any isomorphic quasitriangular Lie bialgebras (see \cite{BoHaReSe2011}).

The isomorphic Lie algebras $\mathfrak{o}(2,1)$, $\mathfrak{sl}(2;\mathbb{R})$, $\mathfrak{su}(1,1)$ and their quantum deformations play very important role in physics as well as in mathematical considerations, so the structure of these deformations should be understood with full clarity. The $\mathfrak{o}(2,1)$ Lie algebra has been used as $D=1$ conformal algebra describing basic symmetries in conformal classical and quantum mechanics \cite{AlFuFu1976}; in such a case $\mathfrak{o}(2,1)$ algebra is realized as a nonlinear realization on the one-dimensional time axis \cite{IvKrLe1989,FeIvLu2011} and can be extended to $\mathfrak{osp}(1|2)$ describing $D=1$ $N=2$ supersymmetric conformal algebra \cite{BoLuTo2003}. In field-theoretic framework the $\mathfrak{o}(2,1)$ Lie algebra describes Lorentz symmetries of three-dimensional relativistic systems with planar $d=2$ space sector, which are often discussed as simplified version of  the  four-dimensional relativistic case. Due to the isomorphism $\mathfrak{o}(2,2)\simeq\mathfrak{o}(2,1)\oplus\mathfrak{o}(2,1)$  our results can be also applied to the description of $D=3$ AdS symmetries \cite{AchTo1986}. We recall that $\mathfrak{o}(2,2)$ symmetry  has been employed in Chern-Simons formulation of $D=3$ gravity \cite{Wi1988, Ca1998, MeSch2003}, with Lorentzian signature and nonvanishing negative cosmological constant. Subsequently, the quantum deformations of $D=3$ Chern-Simons theory have been used for the description of $D=3$ quantum gravity as deformed $D=3$ topological QFT \cite{MeSch2009, BaHeMu2014}. Three-dimensional deformed space-time geometry is also a basis of  historical Ponzano-Regge formulation of $D=3$ quantum gravity \cite{PoRe1968}, which was further developed into spin foam \cite{Li2001} and causal triangulation \cite{ArJu2001} approaches. 

In mathematics  and mathematical physics the importance of $\mathfrak{o}(2,1)$ and its deformations follows also from the unique role of the ${\mathfrak{o}}(2,1)$ algebra as the lowest-dimensional rank one noncompact simple Lie algebra, endowed only with unitary infinite-dimensional representations. One can point out that the programm of deformations of infinite-dimensional modules of quantum-deformed $U({\mathfrak{su}}(1,1))$ algebra has been initiated already more than twenty years ago (see e.g. \cite{MaMiNaNoUe1990}). The $(2+1)$-dimensional models are also important in the theory of classical and quantum integrable systems \cite{Lax1968,Bl1998} with  their symmetries described by Poisson-Lie groups in classical case and after quantization by quantum groups. In particular recently, using  sigma model formulation of (super)string actions (see e.g. \cite{MeTs1998}), there were introduced the integrable deformations of string target (super)spaces obtained by Yang-Baxter deformations \cite{Kl2002}--\cite{ArFrHoRoTs2016} of the principal as well as coset sigma models with symmetries, which may contain $AdS_2\simeq \mathfrak{o}(2;1)$ and $AdS_3\simeq\mathfrak{o}(2,2)$ factors \cite{Ho2015}--\cite{ChLu2006}.

The plan of our paper is the following. In Sect.~2 we consider the complex Lie algebra $\mathfrak{o}(3;\mathbb{C})$ and its all real forms: $\mathfrak{o}(3)\simeq\mathfrak{su}(2)$ and $\mathfrak{o}(2,1)\simeq \mathfrak{su}(1,1)\simeq\mathfrak{sl}(2;\mathbb{R})$. In Sect.~3 we classify all classical $r$-matrices for these real forms and in Sect.~4 we provide the explicite isomorphisms between the real $\mathfrak{su}(1,1)$, $\mathfrak{sl}(2;\mathbb{R})$ and $\mathfrak{o}(2,1)$ bialgebras. In Sect.~5 all three Hopf-algebraic quantizations (explicite quantum deformations) of the real $D=3$ Lorentz symmetry are presented in detail: quantized bases, coproducts and universal $R$-matrices are given. In Sect.~6 we present short summary and outlook. 

\setcounter{equation}{0}
\section{Complex $D=3$ Euclidean Lie algebra $\mathfrak{o}(3;\mathbb{C})$ and its real forms}
We first remind different most popular bases of the complex $D=3$ Euclidean Lie algebra $\mathfrak{o}(3;\mathbb{C})$: {\it metric, Cartesian} and {\it Cartan-Weyl} bases (see \cite{BoLuTo2016}).

The {\it metric} basis contains in its commutation relations an explicite metric, namely, the complex $D=3$ Euclidean Lie algebra $\mathfrak{o}(3;\mathbb{C})$ is generated by three Euclidean basis elements $L_{ij}=-L_{ji}\in\mathfrak{o}(3;\mathbb{C})$ ($i,j=1,2,3$) satisfying the relations
\begin{eqnarray}\label{pr1}
\begin{array}{rcl}
[L_{ij},\,L_{kl}]\!\!& =\!\!&g_{jk}\,L_{il}-g_{jl}\,L_{ik}+g_{il}\,L_{jk}-g_{ik}\,L_{jl},
\end{array}%
\end{eqnarray}
where $g_{ij}$ is the Euclidean metric: $g_{ij}=\mathop{\rm diag}\,(1,1,1)$. The Euclidean algebra $\mathfrak{o}(3;\mathbb{C})$, as a linear space, is a linear envelope of the basis $\{L_{ij}\}$ over $\mathbb{C}$.

The {\it Cartesian} (or {\it physical}) basis of $\mathfrak{o}(3;\mathbb{C})$ is related with the generators $L_{ij}$ as follows
\begin{eqnarray}\label{pr2}
I_{i}\!\!&:=\!\!&-\frac{1}{2}\varepsilon_{ijk}L_{jk}\quad(i,j,k=1,2,3).
\end{eqnarray}
From (\ref{pr1}) and (\ref{pr2}) we get
\begin{eqnarray}\label{pr3}
\begin{array}{rcl}
[I_{i},\,I_{j}]\!\!&=\!\!&\varepsilon_{ijk}I_{k}.
\end{array}%
\end{eqnarray}
If we consider a Lie algebra over $\mathbb{R}$ with the commutation relations (\ref{pr3})  then we get the compact real form $\mathfrak{o}(3):=\mathfrak{o}(3;\mathbb{R})$ with the anti-Hermitian basis
\begin{eqnarray}\label{pr4}
I^{*}_{i}\!\!&=\!\!&-I_{i}\quad(i=1,2,3)\quad{\rm{for}}\;\;\mathfrak{o}(3).
\end{eqnarray}
The real form $\mathfrak{o}(2,1)$ is given by the formulas:
\begin{eqnarray}\label{pr5}
I_{i}^{\dag}\!\!&=\!\!&(-1)^{i-1}I_{i}^{}\quad(i=1,2,3)\quad{\rm{for}}\;\;\mathfrak{o}(2,1).
\end{eqnarray}
For the description of quantum deformations and in particular for the classification of classical $r$-matrices of the complex Euclidean algebra $\mathfrak{o}(3;\mathbb{C})$ and its real forms $\mathfrak{o}(3)$ and $\mathfrak{o}(2,1)$ it is convenient to use the {\it Cartan--Weyl} (CW) basis of the isomorphic complex Lie algebra $\mathfrak{sl}(2;\mathbb{C})$ and its real forms $\mathfrak{su}(2)$, $\mathfrak{sl}(1,1)$ and $\mathfrak{sl}(2,\mathbb{R})$. In the case of $\mathfrak{o}(3)$ the $\mathfrak{su}(2)$ Cartan--Weyl basis can be chosen as follows  
\begin{eqnarray}
&&\begin{array}{rcl}\label{pr6}
&&H\,:=\,\imath I_{3},\qquad\qquad E_{\pm}\,:=\,\imath I_{1}\mp I_{2}, 
\\[4pt]
&&[H,E_{\pm}]\,=\,\pm E_{\pm},\quad[E_{+},E_{-}]\,=\,2H,
\\[4pt]
&&H^{*}\,=\,H,\qquad\qquad\;E_{\pm}^{*}\,=\,E_{\mp}^{},
\end{array}%
\end{eqnarray} 
where the conjugation ($^*$) is the same as in (\ref{pr4})\footnote{The basis elements $E_{\pm}^{},H$ over $\mathbb{C}$ with the defining relations in the second line of (\ref{pr6}) generates the complex Lie algebra $\mathfrak{sl}(2;\mathbb{C})$. The relations in the first line of (\ref{pr6}) reproduce the isomorphism between $\mathfrak{o}(3;\mathbb{C})$ and $\mathfrak{sl}(2;\mathbb{C})$.}.

For the real form $\mathfrak{o}(2,1)$ we will use two CW bases of $\mathfrak{sl}(2;\mathbb{C})$ real forms: $\mathfrak{sl}(1,1)$ and $\mathfrak{sl}(2,\mathbb{R})$. Such bases are given by
\begin{eqnarray}
&&\begin{array}{rcl}\label{pr7}
&&H\,:=\,\imath I_{2},\qquad\qquad E_{\pm}^{}\,:=\,\imath I_{1}\pm I_{3},
\\[4pt]
&&[H,E_{\pm}^{}]\,=\,\pm E_{\pm}^{},\quad[E_{+}^{},E_{-}^{}]\,=\,2H
\end{array}\quad{\rm{for}}\;\;\mathfrak{su}(1,1),%
\\[4pt]
&&\begin{array}{rcl}\label{pr8}
&&H'\,:=\,\imath I_{3},\qquad\qquad E_{\pm}'\,:=\,\imath I_{1}\mp I_{2}, 
\\[4pt]
&&[H',E_{\pm}']\,=\,\pm E_{\pm}',\quad[E_{+}',E_{-}']\,=\,2H'
\end{array}\quad{\rm{for}}\;\;\mathfrak{sl}(2,\mathbb{R}).%
\end{eqnarray}
Both bases $\{E_{\pm},H\}$ and $\{E_{\pm}',H'\}$ have the same commutation relations but they have different reality properties, namely
\begin{eqnarray}
&&\begin{array}{rcccl}\label{pr9}
{H}^{\dag}\!\!&=\!\!&H,\qquad\;\; E_{\pm}^{\dag}\!\!&=\!\!&-E_{\mp}^{}
\end{array}\quad{\rm{for}}\;\;\mathfrak{su}(1,1),
\\[3pt]
&&\begin{array}{rcccl}\label{pr10}
{H'}^{\dag}\!\!&=\!\!&-H',\quad{E_{\pm}'}^{\dag}\!\!&=\!\!&-E_{\pm}'
\end{array}\quad{\rm{for}}\;\;\mathfrak{sl}(2;\mathbb{R}),
\end{eqnarray}
where the conjugation ($^{\dag}$) is the same as in (\ref{pr5})\footnote{It should be noted that in the case of $\mathfrak{su}(1,1)$ the Cartan generator $H$ is compact while for the case $\mathfrak{su}(2,\mathbb{R})$ the generator $H'$ is noncompact.}. The relations between the $\mathfrak{su}(1,1)$ and $\mathfrak{su}(2,\mathbb{R})$ bases look as follows
\begin{eqnarray}
\begin{array}{rcl}\label{pr11}
H\!\!&=\!\!&\displaystyle-\frac{\imath}{2}\big(E_{+}'-E_{-}'\big),
\\[7pt]
E_{\pm}\!\!&=\!\!&\displaystyle\mp\imath H'+\frac{1}{2}\big(E_{+}'+E_{-}'\big).
\end{array}
\end{eqnarray}
\setcounter{equation}{0}
\section{Classical $r$-matrices of $\mathfrak{sl}(2;\mathbb{C})$ and its real forms: $\mathfrak{su}(2)$, $\mathfrak{su}(1,1)$ and $\mathfrak{sl}(2;\mathbb{R})$}

By definition any classical $r$-matrix of arbitrary complex or real Lie algebra $\mathfrak{g}$, $r\in\mathfrak{g}\wedge\mathfrak{g}$, satisfy the classical Yang-Baxter equation (CYBE):
\begin{eqnarray}\label{rm1}
[[r,\,r]]\!\!&=\!\!&\bar{\Omega}~.
\end{eqnarray}
Here $[[\cdot,\cdot]]$ is the Schouten bracket which for any monomial skew-symmetric two-tensors $r_{1}^{}=x\wedge y$ and $r_{2}^{}=u\wedge v$ ($x,y,u,v\in\mathfrak{g}$) is given by\footnote{For general polynomial (a sum of monomials) two-tensors $r_{1}$ and $r_{2}$ one can use the bilinearity of the Schouten bracket.}
\begin{eqnarray}\label{rm2}
\begin{array}{rcl}
[[x\wedge y,\,u\wedge v]]\!\!&:=\!\!&x\wedge\bigl([y,u]\wedge{v}+u\wedge[y,v]\bigr)
\\[4pt]
&&-y\wedge\bigl([x,u]\wedge{v}+u\wedge[x,v]\bigr)
\\[5pt]
\!\!&\phantom{:}=\!\!&[[u\wedge v,\,x\wedge y]]
\end{array}
\end{eqnarray}
and $\bar{\Omega}$ is the $\mathfrak{g}$-invariant element which in the case of $\mathfrak{g}:=\mathfrak{sl}(2;\mathbb{C})$ looks as follows:
\begin{eqnarray}\label{rm3}
\bar{\Omega}\!\!&=\!\!&{\gamma}\,\Omega(\mathfrak{sl}(2;\mathbb{C}))\;=\;\gamma\,(4E_{-}\wedge H\wedge E_{+})
\end{eqnarray}
where $\gamma\in\mathbb{C}$, and $E_{\pm}^{},H$ is the CW basis of $\mathfrak{sl}(2;\mathbb{C})$ with the defining relations on the second line of (\ref{pr6}). 

Firstly we show that {\it any two-tensor of $\mathfrak{sl}(2;\mathbb{C})\wedge\mathfrak{sl}(2;\mathbb{C})$ is a classical $\mathfrak{sl}(2;\mathbb{C})$ $r$-matrix}. Indeed, let 
\begin{eqnarray}\label{rm4}
r\!\!&:=\!\!&\beta_{+}^{}r_{+}^{}+\beta_{0}^{}r_{0}^{}+\beta_{-}^{}r_{-}^{}\quad(\beta_{+}^{},\beta_{0}^{},\beta_{-}^{}\in\mathbb{C})
\end{eqnarray}
be an arbitrary element of $\mathfrak{sl}(2;\mathbb{C})\wedge\mathfrak{sl}(2;\mathbb{C})$, 
where 
\begin{eqnarray}\label{rm5}
r_{+}^{}\,:=\,E_{+}^{}\wedge H,\quad r_{0}^{}\,:=\,E_{+}\wedge E_{-},\quad r_{-}^{}\,:=\,H\wedge E_{-}
\end{eqnarray}
are the basis elements of $\mathfrak{sl}(2;\mathbb{C})\wedge\mathfrak{sl}(2;\mathbb{C})$. Because all terms (\ref{rm5}) are classical $r$-matrices, moreover $[[r_{\pm}^{},r_{\pm}^{}]]=0$, as well as the Schouten brackets of the elemets $r_{\pm}^{}$ with $r_{0}^{}$ are also equal to zero, $[[r_{\pm},r_{0}^{}]]=0$, and we have 
\begin{eqnarray}\label{rm6}
\begin{array}{rcl}
[[r,r]]\!\!&=\!\!&2\beta_{+}^{}\beta_{-}^{}[[r_{+}^{},r_{-}^{}]]+\beta_{0}^{2}[[r_{0}^{},r_{0}^{}]]
\\[6pt]
\!\!&=\!\!&(\beta_{0}^{2}+\beta_{+}^{}\beta_{-}^{})\,(4E_{-}\wedge H\wedge E_{+})\;\equiv\;\gamma\Omega.
\end{array}
\end{eqnarray}
Thus an arbitrary element (\ref{rm4}) is a classical $r$-matrix, and if its coefficients $\beta_{\pm}^{}$, $\beta_{0}^{}$ satisfy the condition  $\gamma:=\beta_{0}^{2}+\beta_{+}^{}\beta_{-}^{}=0$ then it satisfies the homogeneous CYBE, if $\gamma:=\beta_{0}^{2}+\beta_{+}^{}\beta_{-}^{}\neq0$ it satisfies the non-homogeneous CYBE. 

We shall call the parameter $\gamma=\beta_{0}^{2}+\beta_{+}^{}\beta_{-}^{}$ in (\ref{rm6}) the $\gamma$-characteristic of the classical $r$-matrix (\ref{rm4}). It is evident that the $\gamma$-characteristic of the classical $r$-matrix $r$ is invariant under the $\mathfrak{sl}(2;\mathbb{C})$-automorphisms, i.e. any two $r$-matrices $r$ and $r'$, which are connected by a $\mathfrak{sl}(2;\mathbb{C})$-automorphism, have the same $\gamma$-characteristic, $\gamma=\gamma'$. We can show also that any two $\mathfrak{sl}(2;\mathbb{C})$ $r$-matrices $r$ and $r'$ with the same $\gamma$-characteristic can be connected by a $\mathfrak{sl}(2;\mathbb{C})$-automorphism.

There are two types of explicite $\mathfrak{sl}(2;\mathbb{C})$-automorphisms which were presented in \cite{BoLuTo2016}. First type connecting the classical $r$-matrices with zero $\gamma$-characteristic is given by the formulas (see (3.15) in \cite{BoLuTo2016})\footnote{The formulas (\ref{rm7}) are obtained from (3.15) in \cite{BoLuTo2016} by the substitution: $\beta_{0}^{}/(k\beta_{+}\!-\beta_{-}) =-2\tilde{\beta}_{0}$, $\beta_{\pm}^{}/(k\beta_{+}\!-\beta_{-})=\tilde{\beta}_{\pm}^{}$.}:
\begin{eqnarray}\label{rm7}
\begin{array}{rcl}
\varphi_{0}^{}(E_{+})\!\!&=\!\!&\chi(\tilde{\beta}_{+}E_{+}-2\tilde{\beta}_{0}\,H+\tilde{\beta}_{-}E_{-}),
\\[4pt]
\varphi_{0}^{}(E_{-})\!\!&=\!\!&\chi^{-1}(\tilde{\beta}_{-}E_{+}-2\kappa\tilde{\beta}_{0}\,H+\tilde{\beta}_{+}E_{-}),
\\[4pt]
\varphi_{0}^{}(H)\!\!&=\!\!&\tilde{\beta}_{0}\,E_{+}+(\kappa\tilde{\beta}_{+}+\tilde{\beta}_{-})\,H+\kappa\tilde{\beta}_{0}\,E_{-},
\end{array}
\end{eqnarray}
where $\chi$ is a non-zero rescaling parameter (including $\chi=1$), $\kappa$ takes two values $+1$ or $-1$, and the parameters $\tilde{\beta}_{i}$ ($i=+,0,-$) satisfy the conditions:
\begin{eqnarray}\label{rm8}
\tilde{\beta}_{0}^{2}+\tilde{\beta}_{+}^{}\tilde{\beta}_{-}^{}\;=\;0,\quad\;\;\kappa\tilde{\beta}_{+}^{}-\tilde{\beta}_{-}^{}=1.
\end{eqnarray}
Let us consider two general $r$-matrices with zero $\gamma$-characteristics:
\begin{eqnarray}\label{rm9}
\begin{array}{rcl}
r\!\!&:=\!\!&\beta_{+}^{}E_{+}^{}\wedge H+\beta_{0}^{}E_{+}\wedge E_{-}^{}+\beta_{-}^{}H\wedge E_{-}, 
\\[4pt]
r'\!\!&:=\!\!&\beta_{+}'E_{+}\wedge H+\beta_{0}'E_{+}\wedge E_{-}^{}+\beta_{-}'H\wedge E_{-}, 
\end{array}
\end{eqnarray} 
where $\beta_{0}^{2}+\beta_{+}\beta_{-}=0$ and $\beta_{0}^{'2}+\beta_{+}'\beta_{-}'=0$. Moreover, we suppose that the parameters $\beta_{\pm}$ and $\beta_{\pm}'$ satisfy the additional relations:
\begin{eqnarray}\label{rm10}
\kappa\beta_{+}^{}-\beta_{-}^{}=\chi\beta_{+}'-\chi^{-1}\kappa\beta_{-}'\neq\,0,
\end{eqnarray}
where the parameters $\kappa$ and $\chi$ are the same as in (\ref{rm7}). One can check that the following formula is valid:
\begin{eqnarray}\label{rm11}
\begin{array}{rcl}
&&\beta_{+}^{}E_{+}\wedge H+\beta_{0}^{}E_{+}\wedge E_{-}^{}+\beta_{-}^{}H\wedge E_{-}
\\[5pt]
&&=\beta_{+}'\varphi_{0}^{}(E_{+}^{})\wedge\varphi_{0}^{}(H)+\beta_{0}'\varphi_{0}^{}(E_{+}^{})\wedge\varphi_{0}^{}(E_{-})
\\[4pt]
&&\;\;\;+\beta_{-}'\varphi_{0}^{}(H)\wedge\varphi_{0}^{}(E_{-}),
\end{array}
\end{eqnarray} 
where $\varphi_{0 }^{}$ is the $\mathfrak{sl}(2;\mathbb{C})$-automorphism (\ref{rm7}) with the following parameters:
\begin{eqnarray}\label{rm12}
\begin{array}{rcl}
\tilde{\beta}_{0}^{}\!\!&=\!\!&\displaystyle\frac{\beta_{0}^{}(\chi\beta_{+}'+\chi^{-1}\kappa\beta_{-}')-\beta_{0}'(\kappa\beta_{+}^{}+\beta_{-})}{(\kappa\beta_{+}^{}-\beta_{-}^{})(\chi\beta_{+}'-\chi^{-1}\kappa\beta_{-}')}~,
\\[12pt]
\tilde{\beta}_{+}^{}\!\!&=\!\!&\displaystyle\frac{\kappa(\kappa\beta_{+}^{}+\beta_{-}^{})(\chi\beta_{+}'+\chi^{-1}\kappa\beta_{-}')+4\beta_{0}^{}\beta_{0}'}{2(\kappa\beta_{+}^{}-\beta_{-}^{})(\chi\beta_{+}'-\chi^{-1}\kappa\beta_{-}')}+\frac{\kappa}{2}~,
\\[12pt]
\tilde{\beta}_{-}^{}\!\!&=\!\!&\displaystyle\frac{(\kappa\beta_{+}^{}+\beta_{-}^{})(\chi\beta_{+}'+\chi^{-1}\kappa\beta_{-}')+4\kappa\beta_{0}^{}\beta_{0}'}{2(\kappa\beta_{+}^{}-\beta_{-}^{})(\chi\beta_{+}'-\chi^{-1}\kappa\beta_{-}')}-\frac{1}{2}~.
\end{array}
\end{eqnarray}
It is easy to check that as expected the formulas (\ref{rm12}) satisfy the conditions (\ref{rm8}). 

Let us assume in (\ref{rm9}), (\ref{rm11}) and (\ref{rm12}) that the parameters $\beta_{0}'$ and $\beta_{-}'$ are equal to zero. Then the general classical $r$-matrix $r$ in (\ref{rm9}), satisfying the homogeneous CYBE, is reduced to usual Jordanian form by the authomorphism (\ref{rm7}) with the parameters: 
\begin{eqnarray}\label{rm13}
\tilde{\beta}_{0}^{}\;=\;\frac{\beta_{0}^{}}{\kappa\beta_{+}^{}-\beta_{-}^{}}~,\quad   
\tilde{\beta}_{\pm}^{}\;=\;\frac{\beta_{\pm}^{}}{\kappa\beta_{+}^{}-\beta_{-}^{}}~.
\end{eqnarray} 

Second type of $\mathfrak{sl}(2;\mathbb{C})$-automorphism connecting the classical $r$-matrices with non-zero $\gamma$-characteristic is given as follows\footnote{The formulas (\ref{rm8}) are obtained from (3.14) in \cite{BoLuTo2016} by the substitution: $\beta_{0}^{}=2\tilde{\beta}_{0}$, $\beta_{\pm}^{}=-\tilde{\beta}_{\pm}^{}$, $D=4$.}  
\begin{eqnarray}\label{rm14}
\begin{array}{rcl}
\varphi_{1}^{}(E_{+})\!\!&=\!\!&\displaystyle\frac{\chi}{2}
\Bigl((\tilde{\beta}_{0}+1)\,E_{+}+2\tilde{\beta}_{-}H-\frac{\tilde{\beta}_{-}^{2}}{\tilde{\beta}_{0}+1}\,E_{-}\Bigr),
\\[12pt]
\varphi_{1}^{}(E_{-})\!\!&=\!\!&\displaystyle\frac{\chi^{-1}}{2}
\Bigl(\frac{-\tilde{\beta}_{+}^{2}}{\tilde{\beta}_{0}+1}\,E_{+}+2\tilde{\beta}_{+}H+(\tilde{\beta}_{0}+1)E_{-}\Bigr),
\\[12pt]
\varphi_{1}^{}(H)\!\!&=\!\!&\displaystyle\frac{1}{2}\bigl(-\tilde{\beta}_{+}E_{+}+2\tilde{\beta}_{0}H-\tilde{\beta}_{-}E_{-}\bigr),
\end{array}
\end{eqnarray}
where $\chi$ is a non-zero rescaling parameter, and $\tilde{\beta}_{0}^{2}+\tilde{\beta}_{+}\tilde{\beta}_{-}=1$.

Let us consider two general $r$-matrices with non-zero $\gamma$-characteristics:
\begin{eqnarray}\label{rm15}
\begin{array}{rcl}
r\!\!&:=\!\!&\beta_{+}^{}E_{+}^{}\wedge H+\beta_{0}^{}E_{+}\wedge E_{-}^{}+\beta_{-}^{}H\wedge E_{-}, 
\\[4pt]
r'\!\!&:=\!\!&\beta_{+}'E_{+}\wedge H+\beta_{0}'E_{+}\wedge E_{-}^{}+\beta_{-}'H\wedge E_{-}, 
\end{array}
\end{eqnarray} 
where the parameters $\beta_{\pm}^{}$, $\beta_{0}^{}$ and $\beta_{\pm}'$, $\beta_{0}'$ can be equal to zero provided that $\gamma=\beta_{0}^2+\beta_{+}^{}\beta_{-}^{}=\gamma'=(\beta_{0}')^2+\beta_{+}'\beta_{-}'\neq0$, i.e. both $r$-matrices $r$ and $r'$ have the same non-zero $\gamma$-characteristic $\gamma=\gamma'\neq0$. One can check the following relation:
\begin{eqnarray}\label{rm16}
\begin{array}{rcl}
&&\beta_{+}^{}E_{+}\wedge H+\beta_{0}^{}E_{+}\wedge E_{-}^{}+\beta_{-}^{}H\wedge E_{-}
\\[5pt]
&&=\beta_{+}'\varphi_{1}^{}(E_{+}^{})\wedge\varphi_{1}^{}(H)+\beta_{0}'\varphi_{1}^{}(E_{+}^{})\wedge\varphi_{0}^{}(E_{-})
\\[4pt]
&&\;\;\;+\beta_{-}'\varphi_{1}^{}(H)\wedge\varphi_{1}^{}(E_{-}),
\end{array}
\end{eqnarray} 
where $\varphi_{1}^{}$ is the $\mathfrak{sl}(2;\mathbb{C})$-automorphism (\ref{rm14}) with the parameters:
\begin{eqnarray}\label{rm17}
\begin{array}{rcl}
\tilde{\beta}_{0}^{}\!\!&=\!\!&\displaystyle\frac{(\beta_{0}^{}+\beta_{0}')^{2}-(\beta_{+}^{}-\chi\beta_{+}')(\beta_{-}^{}-\chi^{-1}\beta_{-}')}{(\beta_{0}^{}+\beta_{0}')^{2}+(\beta_{+}-\chi\beta_{+}')(\beta_{-}-\chi^{-1}\beta_{-}')}~,
\\[12pt]
\tilde{\beta}_{\pm}^{}\!\!&=\!\!&\displaystyle\frac{2(\beta_{0}^{}+\beta_{0}')(\beta_{\pm}^{}-\chi^{\pm1}\beta_{\pm}')}{(\beta_{0}^{}+\beta_{0}')^{2}+(\beta_{+}^{}-\chi\beta_{+}')(\beta_{-}^{}-\chi^{-1}\beta_{-}')}~.
\end{array}
\end{eqnarray}
It is easy to check that the formulas (\ref{rm17}) satisfy the condition $\tilde{\beta}_{0}^2+\tilde{\beta}_{+}^{}\tilde{\beta}_{-}^{}=1$. 

If we assume in (\ref{rm15})--(\ref{rm17}) that the parameters $\beta_{\pm}'$ are equal to zero then the general classical $r$-matrix $r$ in (\ref{rm15}), satisfying the non-homogeneous CYBE, is reduced to the usual standard form by the automorphism (\ref{rm14}) with the following parameters: 
\begin{eqnarray}\label{rm18}
\tilde{\beta}_{0}^{}\;=\;\frac{\beta_{0}^{}}{\beta_{0}'}~,\quad   
\tilde{\beta}_{\pm}^{}\;=\;\frac{\beta_{\pm}^{}}{\beta_{0}'}~.
\end{eqnarray} 
Finally for $\mathfrak{sl}(2,\mathbb{C})$ we obtain the well-known result:

\textit{For the complex Lie algebra $\mathfrak{sl}(2,\mathbb{C})$ there exists up to $\mathfrak{sl}(2,\mathbb{C})$ automorphisms two solutions of CYBE, namely Jordanian $r_{J}^{}$ and standard $r_{st}^{}$:
\begin{eqnarray}\label{rm19}
r_{J}^{}\!\!&=\!\!&\beta E_{+}\wedge H,\quad[[r_{J}^{},r_{J}^{}]]\;=\;0,
\\[3pt]\label{rm20}
r_{st}^{}\!\!&=\!\!&\beta' E_{+}^{}\wedge E_{-}^{},\quad[[r_{st}^{},r_{st}^{}]]\;=\;{\beta'}^{2}\Omega,
\end{eqnarray}
where the complex parameter $\beta$ in (\ref{rm19}) can be removed by the rescaling automorphism: $\varphi(E_{+})=\beta^{-1}E_{+}$, $\varphi(E_{-})=\beta E_{-}$, $\varphi(H)=H$; in (\ref{rm20}) the parameter $\beta'=e^{\imath\phi}|\beta'|$ for $|\phi|\leq\frac{\pi}{2}$ is effective}.

The general non-reduced expression (\ref{rm4}) is convenient for the application of reality conditions: 
\begin{eqnarray}\label{rm21}
r^{\divideontimes}\!\!&:=\!\!&\beta_{+}^{*}E_{+}^{\divideontimes}\wedge H^{\divideontimes}+\beta_{0}^{*}E_{+}^{\divideontimes}\wedge E_{-}^{\divideontimes}+\beta_{-}^{*}H^{\divideontimes}\wedge E_{-}^{\divideontimes}\,=\,-r,
\end{eqnarray}
where $\divideontimes$ is the conjugation associated with corresponding real form ($\divideontimes=*,\dag$), and $\beta_{i}^{*}$ ($i=+,0,-$) means the complex conjugation of the number $\beta_{i}^{}$. It should be noted that for any classical $r$-matrix $r$, $r^{\divideontimes}$ is again a classical $r$-matrix. Moreover, if $r$-matrix is \textit{anti-real} (anti-Hermitian)\footnote{The anti-real property $r^{\divideontimes}=-r$ is a direct consequence of the reality condition for the co-bracket $\delta(x):=[x\otimes1+1\otimes x,r]$, namely $\delta(x)^{\divideontimes}=\delta(x^{\divideontimes})$ for $\forall x\in\mathfrak{g}^{\divideontimes}(=\{\mathfrak{su}(2),\;\mathfrak{su}(1,1),\;\mathfrak{sl}(2,\mathbb{R})\})$.}, i.e. it satisfies the condition (\ref{rm21}), then its $\gamma$-characteristic is real. Indeed, applying the conjugation $\divideontimes$ to the relation (\ref{rm6}) we have for the left-side: $[[r,r]]^{\divideontimes}=-[[r^{\divideontimes}, r^{\divideontimes}]]=-[[r,r]]$ and for the right-side: $(\gamma\Omega)^{\divideontimes}=-\gamma^{*}\Omega$ for all real forms $\mathfrak{su}(2)$, $\mathfrak{su}(1,1)$, $\mathfrak{su}(2;\mathbb{R})$. It follows that the parameter $\gamma$ is real, $\gamma^{*}=\gamma$. 

I. {\textit{\large The compact real form}} $\mathfrak{su}(2)$ ($H^{*}=H$, $E_{\pm}^{*}=E_{\mp}$).\\ 
In this case it follows from (\ref{rm21}) that
\begin{eqnarray}\label{rm22}
\beta_{0}^{*}\;=\;\beta_{0}^{},\quad\beta_{\pm}^{*}\;=\;\beta_{\mp}^{}.
\end{eqnarray}
If in (\ref{rm4}) $\gamma=\beta_{0}^{2}+\beta_{+}^{}\beta_{-}^{}=0$ then $\beta_{0}^{}\beta_{0}^{*}+\beta_{+}^{}\beta_{+}^{*}=0$ and it follows that $\beta_{0}=\beta_{+}^{}=\beta_{-}^{}=0$, i.e. \textit{any classical $r$-matrix, which satisfies the homogeneous CYBE and the $\mathfrak{su}(2)$ reality condition, is equal zero}. 

If in (\ref{rm4}) $\gamma=\beta_{0}^{2}+\beta_{+}^{}\beta_{-}^{}\neq0$  we have three possible $\mathfrak{su}(2)$ real classical $r$-matrices: 
\begin{eqnarray}\label{rm23}
\begin{array}{rcl}
r_{1}^{}\!\!&:=\!\!&\beta_{0}^{} E_{+}\wedge E_{-}, 
\\[3pt]
r_{2}^{}\!\!&:=\!\!&\beta_{+}^{}E_{+}^{}\wedge H+\beta_{+}^{*}H\wedge E_{-},
\\[3pt]
r_{3}^{}\!\!&:=\!\!&\beta_{+}'E_{+}\wedge H+\beta_{0}'E_{+}\wedge E_{-}^{}+{\beta_{+}'}^{\!\!*}H\wedge E_{-},
\end{array}
\end{eqnarray}
where $\beta_{0}^{}$ and $\beta_{0}'$ are real numbers and we use the conditions (\ref{rm22}). 
The $r$-matrices $r_{i}$ ($i=1,2,3$) satisfy the non-homogeneous CYBE
\begin{eqnarray}\label{rm24}
[[r_{i}^{},r_{i}^{}]]\!\!&=\!\!&\gamma_{i}\Omega,
\end{eqnarray}
where all $\gamma_{i}$ ($i=1,2,3$) are positive: $\gamma_{1}=\beta_{0}^{2}>0$, $\gamma_{2}=\beta_{+}^{}\beta_{+}^{*}>0$, $\gamma_{3}={\beta_{0}'}^{\!2}+\beta_{+}'{\beta_{+}'}^{\!\!*}>0$. 

Let the classical $r$-matrices (\ref{rm15}) be $\mathfrak{su}(2)$-antireal, i.e. their parameters satisfy the reality conditions (\ref{rm22}). It follows that the functions (\ref{rm17}) for $\chi=e^{\imath\phi}$ have the same congugation properties, i.e. $\tilde{\beta}_{0}^{*}=\tilde{\beta}_{0}^{}$, $\tilde{\beta}_{\pm}^{*}=\tilde{\beta}_{\mp}^{}$, and we obtain that the automorphism (\ref{rm14}) with such parameters is $\mathfrak{su}(2)$-real, i.e.:  
\begin{eqnarray}\label{rm25}
\begin{array}{rcl}
\varphi_{1}^{}(E_{\pm})^{*}\!\!&=\!\!&\varphi_{1}^{}(E_{\pm}^{*})\;=\;\varphi_{1}^{}(E_{\mp}),
\\[3pt]
\varphi_{1}^{}(H)^{*}\!\!&=\!\!&\varphi_{1}^{}(H^{*})\;=\;\varphi_{1}^{}(H).
\end{array}
\end{eqnarray}
We see that the $r$-matrices $r_{2}^{}$ and $r_{3}$ in (\ref{rm23}) can be reduced to the standard $r$-matrix $r_{st}^{}:= r_{1}^{}$ using the formula (\ref{rm16}). 

It is easy to see that the standard $r$-matrix $r_{st}^{}=r_{1}^{}$ in (\ref{rm23}) effectively depends only on positive values of the parameter $\alpha:=\beta_{0}^{}$. Indeed, we see that
\begin{eqnarray}\label{rm26}
\alpha\varphi(E_{+})\wedge\varphi(E_{-})\!\!&=\!\!&-\alpha E_{+}\wedge E_{-},
\end{eqnarray}
where $\varphi$ is the simple $\mathfrak{su}(2)$ automorphism: $\varphi(E_{\pm})=E_{\mp}$, $\varphi(H)=-H$, i.e. any negative value of parameter $\alpha$ in $r_{st}^{}$ can be replaced by the positive one.

We obtain the following result:\\
\textit{For the compact real form $\mathfrak{su}(2)$ there exists up to the $\mathfrak{su}(2)$ automorphisms only one solution of CYBE and this solution is the usual standard classical $r$-matrix $r_{st}^{}$:
\begin{eqnarray}\label{rm27}
r_{st}^{}\!\!&:=\!\!&\alpha E_{+}\wedge E_{-},\quad 
[[r_{st}^{},r_{st}^{}]]\;=\;\gamma\Omega,
\end{eqnarray}
where the effective parameter $\alpha$ is a positive number, and $\gamma=\alpha^{2}$}.\\
II. {\large\textit{The non-compact real form}} $\mathfrak{su}(1,1)$ ($H^{\dag}=H$, $E_{\pm}^{\dag}=-E_{\mp}$).\\ 
In this case it follows from (\ref{rm21}) that
\begin{eqnarray}\label{rm28}
\beta_{0}^{*}\;=\;\beta_{0}^{},\quad\beta_{\pm}^{*}\;=\;-\beta_{\mp}^{}.
\end{eqnarray}
If $\beta_{0}^{2}+\beta_{+}^{}\beta_{-}^{}=0$ in (\ref{rm4}) then $\beta_{0}^{}\beta_{0}^{*}-\beta_{+}^{}\beta_{+}^{*}=0$, i.e. $\beta_{\pm}^{}=\pm e^{\pm\imath\phi}|\beta_{0}^{}|$, and we have the following $\phi$-family of $\mathfrak{su}(1,1)$ homogeneous CYBE solutions:
\begin{eqnarray}\label{rm29}
r_{\phi}^{}\!\!&:=\!\!&\beta_{0}^{}\Bigl(e^{i\phi}\frac{|\beta_{0}^{}|}{\beta_{0}^{}}\,E_{+}\wedge H+E_{+}\wedge E_{-}-e^{-i\phi}\frac{|\beta_{0}^{}|}{\beta_{0}^{}}\,H\wedge E_{-}\Bigr),
\end{eqnarray}
where $\beta_{0}^{}$ is real. By using the $\mathfrak{su}(1,1)$-real rescaling automorphism $\varphi(E_{\pm})=\left(-\imath e^{i\phi}\frac{|\beta_{0}^{}|} {\beta_{0}^{}}\right)^{\pm1}E_{\pm}$, $\varphi(H)=H$ we can reduce the $\phi$-family (\ref{rm29}) to $r_{qJ}^{}:=\beta_{0}^{}(\imath E_{+}\wedge H+ E_{+}\wedge E_{-}+\imath H\wedge E_{-})$:
\begin{eqnarray}\label{rm30}
\begin{array}{rcl}
r_{\phi}^{}\!\!&=\!\!&\displaystyle\beta_{0}^{}\Bigl(e^{\imath\phi}\frac{|\beta_{0}^{}|}{\beta_{0}^{}}\,E_{+}\wedge H+E_{+}\wedge E_{-}-e^{-\imath\phi}\frac{|\beta_{0}^{}|}{\beta_{0}^{}}\,H\wedge E_{-}\Big)
\\[10pt]
\!\!&=\!\!&\beta_{0}^{}\Bigl(\imath\bigl(\varphi(E_{+})-\varphi(E_{-})\bigr)\wedge\varphi(H)+\varphi(E_{+})\wedge\varphi(E_{-})\Bigr).
\end{array}
\end{eqnarray}
We shall call a $\mathfrak{su}(1,1)$-real $r$-matrix ''quasi-Jordanian'' if it can not be reduced to Jordanian form by a $\mathfrak{su}(1,1)$-real automorphism, but after complexification of $\mathfrak{su}(1,1)$ it can be reduced to Jordanian form by an appropriate complex $\mathfrak{sl}(2,\mathbb{C})$-authomorphism. Thus all $r$-matrices in the $\phi$-family (\ref{rm29}) are quasi-Jordanian and they are connected with each other by the $\mathfrak{su}(1,1)$-real rescaling automorphism. We take $r_{qJ}^{}$ as an representative of the $\phi$-family. It is easy to see that the quasi-Jordanian $r$-matrix $r_{qJ}^{}$ effectively depends only on positive values of the parameter $\beta_{0}^{}$, indeed,
\begin{eqnarray}\label{rm31}
\begin{array}{rcl}
r_{qJ}^{}\!\!&=\!\!&\beta_{0}^{}(E_{+}\wedge H+E_{+}\wedge E_{-}+H\wedge E_{-})
\\[5pt]
\!\!&=\!\!&-\beta_{0}^{}\bigl((\varphi(E_{+})\wedge\varphi(H)+\varphi(E_{+})\wedge\varphi(E_{-})+\varphi(H)\wedge\varphi(E_{-})\bigr),
\end{array}
\end{eqnarray}
where $\varphi$ is the simple $\mathfrak{su}(1,1)$ automorphism $\varphi(E_{\pm})=E_{\mp}$, $\varphi(H)=-H$, i.e. any negative value of parameter $\beta_{0}^{}$ in $r_{qJ}^{}$ can be changed into a positive one.

In the case $\beta_{0}^{2}+\beta_{+}^{}\beta_{-}^{}\neq0$ in (\ref{rm4}) we have four versions of $\mathfrak{su}(1,1)$-real classical $r$-matrices. Two of them are characterized by positive value of $\gamma_{i}$, ($i=1,2$):
\begin{eqnarray}\label{rm32}
\begin{array}{rcl}
&&r_{1}^{}\;:=\;\beta_{0}^{}E_{+}\wedge E_{-},
\\[4pt]
&&r_{2}^{}\;:=\;\beta_{+}'E_{+}^{}\wedge H+\beta_{0}'E_{+}\wedge E_{-}^{}-{\beta_{+}'}^{\!\!*}H\wedge E_{-},
\\[4pt]
&&[[r_{i}^{},r_{i}^{}]]\;:=\;\gamma_{i}\Omega\quad (i=1,2),
\end{array}
\end{eqnarray}
where $\beta_{0}^{}$ and $\beta_{0}'$ are real (see (\ref{rm28})), and $\gamma_{1}=\beta_{0}^{2}>0$, $\gamma_{2}=\beta_{0}'{\beta_{0}'}^{*}-\beta_{+}'{\beta_{+}'}^{*}>0$. The remaining two are with negative values of $\gamma_{i}$, ($i=3,4$):
\begin{eqnarray}\label{rm33}
\begin{array}{rcl}
&&r_{3}^{}\;:=\;\beta_{+}''E_{+}\wedge H-{\beta_{+}''}^{\!*}H\wedge E_{-},
\\[4pt]
&&r_{4}^{}\;:=\;\beta_{+}'''E_{+}\wedge H+\beta_{0}'''E_{+}\wedge E_{-}-{\beta_{+}'''}^{*}H\wedge E_{-}^{},
\\[4pt]
&&[[r_{i}^{},r_{i}^{}]]\;:=\;\gamma_{i}\Omega\quad (i=3,4),
\end{array}
\end{eqnarray}
where $\beta_{0}'''$ is real (see (\ref{rm28})), and $\gamma_{3}=-\beta_{+}''{\beta_{+}''}^{*}<0$, $\gamma_{4}=\beta_{0}'''{\beta_{0}'''}^{*}-\beta_{+}'''{\beta_{+}'''}^{*}<0$.

Let the classical $r$-matrices (\ref{rm15}) be $\mathfrak{su}(1,1)$-antireal, i.e. their parameters satisfy the reality conditions (\ref{rm28}). In such case the functions (\ref{rm17}) for $\chi=e^{\imath\phi}$ have the same conjugation properties, i.e. $\tilde{\beta}_{0}^{*}=\tilde{\beta}_{0}^{}$, $\tilde{\beta}_{\pm}^{*}=-\tilde{\beta}_{\mp}^{}$, and we obtain that the automorphism (\ref{rm14}) with these parameters is $\mathfrak{su}(1,1)$-real, i.e.:  
\begin{eqnarray}\label{rm34}
\begin{array}{rcl}
\varphi_{1}^{}(E_{\pm})^{\dag}\!\!&=\!\!&\varphi_{1}^{}(E_{\pm}^{\dag})\;=\;-\varphi_{1}^{}(E_{\mp}),
\\[4pt]
\varphi_{1}^{}(H)^{\dag}\!\!&=\!\!&\varphi_{1}^{}(H^{\dag})\;=\;\varphi_{1}^{}(H).
\end{array}
\end{eqnarray}
It allows to reduce the $r$-matrix $r_{2}^{}$ to the standard $r$-matrix $r_{st}^{}:=r_{1}^{}$ for $\gamma_{1}^{}=\gamma_{2}^{}>0$ and the $r$-matrix $r_{3}^{}$ to the $r$-matrix $r_{4}^{}$  for $\gamma_{3}^{}=\gamma_{4}^{}<0$ by use of the formula (\ref{rm16}). By analogy to the notation of quasi-Jordanian $r$-matrix we shall call the $r$-matrices $r_{3}$ and $r_{4}$ as quasi-standard ones and take $r_{qst}^{}:=\alpha(E_{+}+E_{-})\wedge H$ as their representative\footnote{The $r$-matrix $r_{qst}^{}$ is connected with $r_{3}^{}$ (\ref{rm33}) by the following way. Substituting $\beta_{+}^{}=|\beta_{+}^{}|e^{\imath\phi}$ in $r_{3}^{}$ (\ref{rm33}) and using the $\mathfrak{su}(1,1)$-real rescaling automorphism $\varphi(E_{\pm})=e^{\pm i\phi}E_{\pm}$, $\varphi(H)=H$ we obtain $r_{qst}^{}$ with $\alpha=|\beta_{+}^{}|$.}.
 
Finally for $\mathfrak{su}(1,1)$ we obtain:\\
\textit{For the non-compact real form $\mathfrak{su}(1,1)$ there exists up to $\mathfrak{su}(1.1)$ automorphisms three solutions of CYBE, namely quasi-Jordanian $r_{qJ}$, standard $r_{st}$ and quasi-standard $r_{qst}$:
\begin{eqnarray}\label{rm35}
r_{qJ}^{}\!\!&=\!\!&\frac{\alpha}{2}\big(\imath(E_{+}-E_{-})\wedge H+E_{+}\wedge E_{-}\big),
\quad[[r_{qJ}^{},r_{qJ}^{}]]\;=\;0,
\\[4pt]\label{rm36}
r_{st}^{}\!\!&=\!\!&\alpha E_{+}\wedge E_{-},\quad[[r_{st}^{},r_{st}^{}]]\;=\;\alpha^{2}\Omega,
\\[4pt]\label{rm37}
r_{qst}^{}\!\!&=\!\!&\alpha(E_{+}\!+E_{-})\wedge H,\quad
[[r_{qst}^{},r_{qst}^{}]]\;=\;-\alpha^{2}\Omega,
\end{eqnarray}
where $\alpha$ effectively is a positive number}.\\
III. \textit{\large The non-compact real form} $\mathfrak{sl}(2;\mathbb{R})$ (${H'}^{\dag}=-H'$, ${E_{\pm}'}^{\dag}=-E_{\pm}'$). \\
In this case from (\ref{rm21}) we obtain
\begin{eqnarray}\label{rm38}
\beta_{0}^{*}\;=\;-\beta_{0},\quad\beta_{\pm}^{*}\;=\;-\beta_{\pm},
\end{eqnarray} 
i.e. all parameters $\beta_{i}^{}$ ($i=+,0,-$) are purely imaginary. 

Consider the case $\beta_{0}^{2}+\beta_{+}^{}\beta_{-}^{}=0$ in (\ref{rm4}). We have three $\mathfrak{su}(2;\mathbb{R})$ solutions of the homogeneous CYBE:
\begin{eqnarray}\label{rm39}
\begin{array}{ccc}
&&r_{1}'\;=\;\beta_{+}^{}E_{+}'\wedge H',\quad r_{2}'\;=\;\beta_{-}^{}H'\wedge E_{-}',
\\[4pt]
&&r_{3}'\;=\;\beta_{+}',E_{+}'\wedge H'+\beta_{0}'E_{+}'\wedge E_{-}'+\beta_{-}'H'\wedge E_{-}',
\end{array}
\end{eqnarray}
where all parameters $\beta_{i}^{}$ ($i=+,-$), $\beta_{i}'$ ($i=+,0,-$) are purely imaginary, and ${\beta_{0}'}^{\!2}+\beta_{+}'\beta_{-}'=0$. 

If the classical $r$-matrices (\ref{rm9}), where all generators $H$, $E_{\pm}^{}$ are replaced by $H'$, $E_{\pm}'$, are $\mathfrak{sl}(2;\mathbb{R})$-antireal, i.e. their parameters satisfy the reality conditions (\ref{rm38}), then for the real parameter $\chi$ all functions (\ref{rm12}) are real, i.e. $\tilde{\beta}_{0}^{*}=\tilde{\beta}_{0}^{}$, $\tilde{\beta}_{\pm}^{*}=\tilde{\beta}_{\pm}^{}$. We obtain that the automorphism of the type (\ref{rm7}) with such parameters is $\mathfrak{sl}(2;\mathbb{R})$-real, i.e.:  
\begin{eqnarray}\label{rm40}
\begin{array}{rcl}
\varphi_{0}^{}(E_{\pm}')^{\dag}\!\!&=\!\!&\varphi_{0}^{}({E_{\pm}'}^{\dag})\;=\;-\varphi_{0}^{}(E_{\pm}'),
\\[4pt]
\varphi_{0}^{}(H')^{\dag}\!\!&=\!\!&\varphi_{0}^{}({H'}^{\dag})\;=\;-\varphi_{0}^{}(H').
\end{array}
\end{eqnarray}
It allows to reduce the $r$-matrices $r_{2}'$ and $r_{3}'$ in (\ref{rm39}) to the Jordanian $r$-matrix $r_{J}':=r_{1}'$ by using the formula (\ref{rm11}).

In the case $\beta_{0}^{2}+\beta_{+}^{}\beta_{-}^{}\neq0$ in (\ref{rm4}) we have seven versions of $\mathfrak{sl}(2;\mathbb{R})$-real classical $r$-matrices. Five of them are with negative values of $\gamma_{i}$, ($i=1,2,\ldots,5$):
\begin{eqnarray}\label{rm41}
\begin{array}{rcl}
&&r_{1}'\;:=\;\beta_{0}^{}E_{+}'\wedge E_{-}',
\\[4pt]
&&r_{2}'\;:=\;\beta_{+}^{}E_{+}'\wedge H'+\beta_{0}^{}E_{+}'\wedge E_{-}',
\\[4pt]
&&r_{3}'\;:=\;\beta_{0}^{}E_{+}'\wedge E_{-}'+\beta_{-}^{}H'\wedge E_{-}',
\\[4pt]
&&r_{4}'\;:=\;\beta_{+}'E_{+}'\wedge H'+\beta_{-}'H'\wedge E_{-}',
\\[4pt]
&&r_{5}'\;:=\;\beta_{+}''E_{+}'\wedge H'+\beta_{0}''E_{+}'\wedge E_{-}'+\beta_{-}''H'\wedge E_{-}',
\\[4pt]
&&[[r_{i}',r_{i}']]\;:=\;\gamma_{i}\Omega'\quad (i=1,2,\ldots,5),
\end{array}
\end{eqnarray}
where all parameters $\beta$ are purely imaginary, and $\gamma_{1}=\gamma_{2}=\gamma_{3}=\beta_{0}^{2}<0$, $\gamma_{4}=\beta_{+}'\beta_{-}'<0$, $\gamma_{5}=\beta_{0}''+\beta_{+}''\beta_{-}''<0$; $\Omega'$ is the $\mathfrak{sl}(2;\mathbb{R})$-invariant element\footnote{Using (\ref{pr11}) it is easy to check that $\Omega'=\Omega$ (see the formula (\ref{rm3})).}: $\Omega'\;=\;\gamma\,(4E_{-}'\wedge H'\wedge E_{+}')$. The remaining two $r$-matrices $r_{i}'$ ($i=6,7$) have positive values of $\gamma_{i}$:
\begin{eqnarray}\label{rm42}
\begin{array}{rcl}
&&r_{6}'\;:=\;\beta_{+}'''E_{+}'\wedge H'+\beta_{-}'''H'\wedge E_{-}',
\\[4pt]
&&r_{7}'\;:=\;\beta_{+}''''E_{+}'\wedge H'+\beta_{0}''''E_{+}'\wedge E_{-}'+\beta_{-}''''^{}H'\wedge E_{-}',
\\[4pt]
&&[[r_{i}',r_{i}']]\;:=\;\gamma_{i}\Omega'\quad (i=6,7),
\end{array}
\end{eqnarray}
where $\gamma_{6}=\beta_{+}'''\beta_{-}'''>0$ and $\gamma_{7}={\beta_{0}''''}^{2}+\beta_{+}''''\beta_{-}''''>0$. 

Let the classical $r$-matrices (\ref{rm15}) be $\mathfrak{sl}(2;\mathbb{R})$-antireal, i.e.  with their parameters satisfying the reality conditions (\ref{rm38}). In such way the functions (\ref{rm17}) for real $\chi$ are real, i.e. $\tilde{\beta}_{0}^{*}=\tilde{\beta}_{0}^{}$, $\tilde{\beta}_{\pm}^{*}=\tilde{\beta}_{\pm}^{}$, and we obtain that the automorphism (\ref{rm14}) with such parameters is $\mathfrak{sl}(2;\mathbb{R})$-real. We can conclude that for the case of the negative $\gamma$-characteristics $\gamma_{i}^{}<0$ ($i=1,\dots,5$) all $r$-matrices $r_{i}^{}$ $(i=2,\ldots,5)$ in (\ref{rm41}) are reduced to the standard formula $r_{st}':=r_{1}'$ and in the case of the positive $\gamma$-characteristics $\gamma_{i}^{}>0$ ($i=6,7$) the classical $r$-matrix $r_{7}'$ in (\ref{rm42}) is reduced to the quasi-standard $r$-matrix $r_{qst}':=r_{6}'$. 

Let us show that the $r$-matrix $r_{qst}'$ effectively depend only on one positive parameter. Indeed, it is easy to see that
\begin{eqnarray}
\begin{array}{rcl}\label{rm43}
r_{qst}'\!\!&=\!\!&\displaystyle\sqrt{\beta_{+}^{}\beta_{-}^{}}
\biggl(\frac{\beta_{+}^{}}{\sqrt{\beta_{+}^{}\beta_{-}^{}}}\;E_{+}'\wedge H'+\frac{\beta_{-}^{}}{\sqrt{\beta_{+}^{}\beta_{-}^{}}}\;H'\wedge E_{-}'\biggr)
\\[15pt]
\!\!&=\!\!&\imath\alpha\big(\varphi(E_{+}')+\varphi(E_{-}')\big)\wedge\varphi(H'),
\end{array}
\end{eqnarray}\\[-10pt]
where $\varphi$ is the $\mathfrak{sl}(2,\mathbb{R})$-real automorphism: $\varphi(E_{\pm}')=\frac{\mp\imath\beta_{\pm}^{}}{\sqrt{\beta_{+}^{}\beta_{-}^{}}}E_{\pm}'$, $\varphi(H')=H'$, and $\alpha=\sqrt{\beta_{+}^{}\beta_{-}^{}}$ is positive.\\
Finally for $\mathfrak{sl}(2,\mathbb{R})$ we obtain the following result:

\textit{For the non-compact real form $\mathfrak{sl}(2,\mathbb{R})$ there exists up to $\mathfrak{sl}(2,\mathbb{R})$ automorphisms three solutions of CYBE, namely Jordanian $r_{J}'$, standard $r_{st}'$ and quasi-standard $r_{qst}'$:
\begin{eqnarray}\label{rm44}
r_{J}'\!\!&=\!\!&\imath\alpha E_{+}'\wedge H',\quad[[r_{J}',r_{J}']]\;=\;0,
\\[3pt]\label{rm45}
r_{st}'\!\!&=\!\!&\imath\alpha E_{+}'\wedge E_{-}',\quad[[r_{st}',r_{st}']]\;=\;-\alpha^{2}\Omega',
\\[3pt]\label{rm46}
r_{qst}'\!\!&=\!\!&\imath\alpha_{}(E_{+}'+E_{-}')\wedge H',\quad[[r_{qst}',r_{qst}']]\;=\;\alpha_{}^{2}\Omega',
\end{eqnarray}
where the parameter $\alpha$ is a positive number}.

\setcounter{equation}{0}
\section{Explicite isomorphism between $\mathfrak{su}(1,1)$ and $\mathfrak{sl}(2;\mathbb{R})$ bialgebras and its application to $\mathfrak{o}(2,1)$ quantizitions}
Using the formulas (\ref{pr7}) and (\ref{pr8}) we express the triplets of the classical $\mathfrak{su}(1,1)$ and $\mathfrak{sl}(2;\mathbb{R})$ $r$-matrices in terms of the $\mathfrak{o}(2,1)$ basis (\ref{pr3}), (\ref{pr5}). We get the following results.\\ 
({\it i}) The $\mathfrak{su}(1,1)$ case:
\begin{eqnarray}
&&\begin{array}{rcl}\label{is1}
\phantom{aaa} r_{qJ}^{}\!\!&=\!\!&\displaystyle\frac{\alpha}{2}\big(\imath(E_{+}^{}-E_{-}^{})\wedge H+E_{+}^{}\wedge E_{-}^{})\big)
\\[5pt]
\!\!&=\!\!&-\alpha(\imath I_{1}^{}-I_{2}^{})\wedge I_{3}^{},\quad[[r_{qJ}^{},r_{qJ}^{}]]\;=\;0,
\end{array}
\\[5pt]
&&\begin{array}{rcl}\label{is2}
&&r_{st}^{}\;=\;\alpha E_{+}^{}\wedge E_{-}^{}\;=\;-2\imath\alpha I_{1}^{}\wedge I_{3}^{},
\\[4pt]
&&[[r_{st}^{},r_{st}^{}]]\;=\;\alpha^{2}\Omega,
\end{array}
\\[5pt]
&&\begin{array}{rcl}\label{is3}
&&r_{qst}^{}\;=\;\alpha(E_{+}^{}+E_{-}^{})\wedge H\;=\;-2\alpha I_{1}\wedge I_{2},
\\[4pt]
&&[[r_{qst}^{},r_{qst}^{}]]\;=\;-\alpha^{2}\Omega,
\end{array}
\end{eqnarray}
where the $\mathfrak{o}(2,1)$-invariant element $\Omega$ expressed in terms of the Cartesian basis (\ref{pr3}) satisfying the reality condition (\ref{pr5}) looks as follows
\begin{eqnarray}\label{is4}
\Omega\!\!&=\!\!&-8I_{1}\wedge I_{2}\wedge I_{3}.
\end{eqnarray} 
({\it ii}) The $\mathfrak{su}(2;\mathbb{R})$ case:
\begin{eqnarray}
&&\begin{array}{rcl}\label{is5}
&&r_{J}'\;=\;\imath\alpha E_{+}'\wedge H'=-\alpha(\imath I_{1}^{}-I_{2}^{})\wedge I_{3}^{}),
\\[5pt]
&&[[r_{J}',r_{J}']]\;=\;0,
\end{array}
\\[5pt]
&&\begin{array}{rcl}\label{is6}
&&r_{st}'\;=\;\imath\alpha E_{+}'\wedge E_{-}'\;=\;-2\alpha I_{1}^{}\wedge I_{2}^{},
\\[5pt]
&&[[r_{st}',r_{st}']]\;=\;-\alpha^{2}\Omega'.
\end{array}
\\[5pt]
&&\begin{array}{rcl}\label{is7}
&&r_{qst}'\;=\;\imath\alpha(E_{+}'+E_{-}')\wedge H'\;=\;-2\imath\alpha I_{1}^{}\wedge I_{3}^{},
\\[5pt]
&&[[r_{qst}',r_{qst}']]\;=\;\alpha^{2}\Omega',
\end{array}
\end{eqnarray} 
where $\Omega'=\Omega$. 

Comparing the $r$-matrix expressions (\ref{is1})--(\ref{is3}) with (\ref{is5})--(\ref{is7}) we obtain that
\begin{eqnarray}\label{is8}
r_{qJ}^{}\!\!&=\!\!&r_{J}'\;=\;-\alpha(\imath I_{1}^{}-I_{2}^{})\wedge I_{3}^{}),
\\ \label{is9}
r_{st}^{}\!\!&=\!\!&r_{qst}'\;=\;-2\imath\alpha I_{1}^{}\wedge I_{3}^{},
\\ \label{is10}
r_{qst}^{}\!\!&=\!\!&r_{st}'\;=\;-2\alpha I_{1}^{}\wedge I_{2}^{},
\end{eqnarray}
We see that the quasi-Jordanian $r$-matrix $r_{qJ}^{}$ in the $\mathfrak{su}(1,1)$ basis is the same as the Jordanian $r$-matrix $r_{J}'$ in the $\mathfrak{sl}(2;\mathbb{R})$ basis, and the standard $r$-matrix $r_{st}^{}$ in the $\mathfrak{su}(1,1)$ basis becomes the quasi-standard $r$-matrix $r_{qst}'$ in the $\mathfrak{sl}(2;\mathbb{R})$ basis. Conversely, the quasi-standard $r$-matrix $r_{qst}^{}$ in the $\mathfrak{su}(1,1)$ basis is the same as the standard $r$-matrix $r_{st}'$ in the $\mathfrak{sl}(2;\mathbb{R})$ basis.

\textit{The relations (\ref{is8})--(\ref{is10}) show that the $\mathfrak{su}(1,1)$ and $\mathfrak{sl}(2;\mathbb{R})$ bialgebras are isomorphic}. This result finally resolves the doubts about isomorphism of these two bialgebras (for example, see \cite{BoHaReSe2011}). 


Using the isomorphisms of the $\mathfrak{su}(1,1)$ and $\mathfrak{sl}(2;\mathbb{R})$ bialgebras we take as basic $r$-matrices for the $D=3$ Lorentz algebra $\mathfrak{o}(2,1)$ the following ones:
\begin{eqnarray}\label{is11}
r_{st}^{}\!\!&=\!\!&-2\imath\alpha I_{1}^{}\wedge I_{3}^{}\;=\;\alpha E_{+}^{}\wedge E_{-}^{}, 
\\ \label{is12} 
r_{st}'\!\!&=\!\!&-2\alpha I_{1}^{}\wedge I_{2}^{}\;=\;\imath\alpha E_{+}'\wedge E_{-}',
\\ \label{is13}
r_{J}'\!\!&=\!\!&-\alpha(\imath I_{1}^{}-I_{2}^{})\wedge I_{3}^{}\;=\;\imath\alpha E_{+}'\wedge H'.
\end{eqnarray}
The first two $r$-matrices $r_{st}^{}$ and $r_{st}'$ with the effective positive parameter $\alpha$ correspond to the $q$-analogs of $\mathfrak{su}(1,1)$ and $\mathfrak{sl}(2;\mathbb{R})$ real algebras, the third $r$-matrix $r_{J}'$ presents the Jordanian twist deformation of $\mathfrak{sl}(2;\mathbb{R})$. In the next section we shall show how to quantize the $r$-matrices (\ref{is11})--(\ref{is13}) in an explicite form.

\setcounter{equation}{0}
\section{Quantizations of the $D=3$ Lorentz symmetry}
The $q$-analogs of the universal enveloping algebras $U(\mathfrak{g})$ for the real Lie algebras $\mathfrak{g}=\mathfrak{su}(1,1)$, $\mathfrak{sl}(2;\mathbb{R})$ were already considered (see e.g. \cite{Ma1995, MaMiNaNoUe1990, KlSch1997}) and they are given as follows. The quantum deformation ($q$-analog) of $U(\mathfrak{g})$ is an unital associative algebra $U_{q}(\mathfrak{g})$ with generators $X_{\pm}$, $q^{\pm X_{0}}$ and the defining relations:
\begin{eqnarray}\label{qu1}
\begin{array}{rcl}
q^{X_{0}}q^{-X_{0}}\!\!&=\!\!&q^{-X_{0}}q^{X_{0}}\;=\;1,
\\[4pt] 
q^{X_{0}}X_{\pm}\!\!&=\!\!&q^{\pm1}X_{\pm}q^{X_{0}},
\\[4pt] 
[X_{+},\,X_{-}]\!\!&=\!\!&\displaystyle\frac{q^{2X_{0}}-q^{-2X_{0}}}{q-q^{-1}}\,,
\end{array}
\end{eqnarray}
with the reality conditions:
\begin{eqnarray}\label{qu2}
\begin{array}{rcl}
&&(i)\;\;\;X_{\pm}^{\dag}\,=\,-X_{\mp}^{},\quad (q^{X_0})^{\dag}\,=\,q^{X_0},\quad q\,:=\,e^{\alpha}\,\quad\,{\rm for}\;\;U_{q}(\mathfrak{su}(1,1)),
\\[5pt]
&&(ii)\;\;X_{\pm}^{\dag}\,=\,-X_{\pm}^{},\quad (q^{X_0})^{\dag}\,=\,q^{X_0},\quad q\,:=\,e^{\imath\alpha}\,\quad{\rm for}\;\;U_{q}(\mathfrak{sl}(2;\mathbb{R})),
\end{array}
\end{eqnarray}
where $\alpha$ is real in accordance with (\ref{is11}) and (\ref{is12}).  

A Hopf structure on $U_{q}(\mathfrak{g})$ ($\mathfrak{g}=\mathfrak{su}(1,1)$, $\mathfrak{sl}(2;\mathbb{R})$) is defined with help of three additional operations: coproduct (comultiplication) $\Delta_{q}$, antipode $S_{q}$ and counit $\epsilon_{q}$:
\begin{eqnarray}\label{qu3}
\begin{array}{rcl}
\Delta_{q}(q^{\pm X_{0}})\!\!&=\!\!&q^{\pm X_{0}}\otimes q^{\pm X_{0}},
\\[4pt] 
\Delta_{q}(X_{\pm}^{})\!\!&=\!\!&X_{\pm}^{}\otimes q^{X_{0}}+q^{-X_{0}}\otimes X_{\pm}^{},
\\[4pt]
S_{q}(q^{\pm X_{0}})\!\!&=\!\!&q^{\mp X_{0}},\quad S_{q}(X_{\pm}^{})\;=\;-q^{\pm1}X_{\pm}^{},
\\[4pt]
\epsilon_{q}(q^{\pm X_{0}})\!\!&=\!\!&1,\qquad\;\;\epsilon_{q}(X_{\pm}^{})\;=\;0,
\end{array}
\end{eqnarray}
with the reality conditions\footnote{$\Delta_{q}^{\dag}(X)\!:=\!(\Delta_{q}(X))^{\dag\otimes\dag}$.}:
\begin{eqnarray}\label{qu4}
\begin{array}{rcl}
\Delta_{q}^{\dag}(X)\,=\,\Delta_{q}(X^{\dag}),\quad S_{q}^{\dag}(X)\,=\,S_{q}^{-1}(X^{\dag}),\quad \epsilon_{q}^{*}(X)\,=\,\epsilon_{q}(X^{\dag}) 
\end{array}
\end{eqnarray}
for any $X\in U_{q}(\mathfrak{g})$. The quantum algebra $U_{q}(\mathfrak{g})$ is endowed also with the {\it opposite} Hopf structure: opposite  coproduct $\tilde{\Delta}_{q}$\footnote{The opposite (transformed) coproduct $\tilde{\Delta}_{q}(\cdot)$ is a coproduct with permuted components, i.e. $\tilde{\Delta}_{q}(\cdot)=\tau\circ\Delta_{q}(\cdot)$ where $\tau$ is the flip operator: $\tau\circ\sum X_{(1)}\otimes X_{(2)}=\sum X_{(2)}\otimes X_{(1)}$.}, corresponding antipode $\tilde{S}_{q}$ and counit $\tilde{\epsilon}_{q}$.

An invertible element $R_{q}:=R_{q}(\mathfrak{g})$ which satisfies the relations: 
\begin{eqnarray}\label{qu5}
\begin{array}{rcl}
R_{q}\Delta_{q}(X)\!\!&=\!\!&\tilde{\Delta}_{q}(X)R_{q},\quad \forall X\in U_{q}(\mathfrak{g}),  
\\[5pt]
(\Delta_{q}\otimes{\rm id})R_{q}\!\!&=\!\!&R_{q}^{13}R_{q}^{23},\quad({\rm id}\otimes\Delta_{q})R_{q}\;=\;R_{q}^{12}R_{q}^{13}
\end{array}
\end{eqnarray}
as well as, due to (\ref{qu5}), the quantum Yang-Baxter equation (QYBE)
\begin{eqnarray}\label{qu6}
\begin{array}{rcl}
R_{q}^{12}R_{q}^{13}R_{q}^{23}\!\!&=\!\!&R_{q}^{23}R_{q}^{13}R_{q}^{12}
\end{array}
\end{eqnarray}
is called the {\it universal $R$-matrix}. Let $U_{q}(\mathfrak{b}_{+})$ and  $U_{q}(\mathfrak{b}_{-})$ be quantum Borel subalgebras of $U_{q}(\mathfrak{g})$, generated by $X_{+}$, $q^{\pm X_{0}}$ and $X_{-}$, $q^{\pm X_{0}}$ respectively. We denote by $T_{q}(\mathfrak{b}_{+}\otimes\mathfrak{b}_{-})$ the Taylor extension of $U_{q}(\mathfrak{b}_{+})\otimes U_{q}(\mathfrak{b}_{-})$\footnote{$T_{q}(\mathfrak{b}_{+}\otimes\mathfrak{b}_{-})$  is an associative algebra generated by formal Taylor series of the monomials $X_{+}^{n}\otimes X_{-}^{m}$ with coefficients which are rational functions of $q^{\pm X_{0}}$, $q^{\pm X_{0}\otimes X_{0}}$, provided that all values $|n-m|$ for each formal series are bounded, $|n-m|<N$.}. One can show (see \cite{KhTo1991,KhTo1992}) that \textit{there exists unique solution of equations (\ref{qu5}) in the space $T_{q}(\mathfrak{b}_{+}\otimes\mathfrak{b}_{-})$ and such solution has the following form}
\begin{eqnarray}\label{qu7}
\begin{array}{rcl}
R_{q}(\mathfrak{g})\;:=\;R_{q}^{\succ}\!\!&=\!\!&\exp_{q^{-2}}\big((q-q^{-1})X_{+}q^{-X_{0}}\otimes q^{X_{0}}X_{-}\big)q^{2X_{0}\otimes X_{0}}
\\[5pt]
\!\!&=\!\!&q^{2X_{0}\otimes X_{0}}\exp_{q^{-2}}\big((q-q^{-1})X_{+}q^{X_{0}}\otimes q^{-X_{0}}X_{-}\big),
\end{array}
\end{eqnarray}
\textit{where} $q\;=\;e^{\alpha}$ for $U_{q}(\mathfrak{su}(1,1))$ and $q=e^{\imath\alpha}$ for $U_{q}(\mathfrak{sl}(2;\mathbb{R}))$. Here we use the standard definition of the $q$-exponential: 
\begin{eqnarray}\label{qu8}
\begin{array}{rcl}
\exp_{q}(x)\!\!&:=\!\!&\displaystyle\sum_{n\geq0}\frac{x^{n}}{(n)_{q}!}\;,\quad (n)_{q}\,:=\,\frac{(1-q^{n})}{(1-q)}~, 
\\[15pt]
(n)_{q}!\!\!&:=\!\!&(1)_{q}(2)_{q}\dots(n)_{q}.
\end{array} 
\end{eqnarray}
Analogously, \textit{there exits unique solution of equations (\ref{qu5}) in the space $T_{q}(\mathfrak{b}_{-}\otimes\mathfrak{b}_{+})=\tau\circ T_{q}(\mathfrak{b}_{+}\otimes\mathfrak{b}_{-})$ and such solution is given by the formula} 
\begin{eqnarray}\label{qu9}
\begin{array}{rcl}
R_{q}(\mathfrak{g})\;:=\;R_{q}^{\prec}\!\!&=\!\!&\exp_{q^{2}}\big((q^{-1}-q)X_{-}q^{-X_{0}}\otimes q^{X_{0}}X_{+}\big)q^{-2X_{0}\otimes X_{0}}
\\[5pt]
\!\!&=\!\!&q^{-2X_{0}\otimes X_{0}}\exp_{q^{2}}\big((q^{-1}-q)X_{-}q^{X_{0}}\otimes q^{-X_{0}}X_{+}\big),
\end{array}
\end{eqnarray}
\textit{where $q$ satisfies the conditions (\ref{qu2})}.\\
As formal Taylor series the solutions (\ref{qu7}) and (\ref{qu9}) are independent and they are related by the relation 
\begin{eqnarray}\label{qu10}
R_{q}^{\prec}\!\!&=\!\!&\tau\circ R_{q^{-1}}^{\succ}.
\end{eqnarray}
It should be noted also that 
\begin{eqnarray}\label{qu11}
(R_{q}^{\succ})^{-1}\!\!&=\!\!&R_{q^{-1}}^{\succ},\quad (R_{q}^{\prec})^{-1}=R_{q^{-1}}^{\prec}.
\end{eqnarray}
From the explicite forms (\ref{qu7}) and (\ref{qu9}) we see that 
\begin{eqnarray}\label{qu12}
\begin{array}{rcl}
&(R_{q}^{\succ})^{\dag}=\tau\circ R_{q}^{\succ}=(R_{q}^{\prec})^{{-1}},\quad(R_{q}^{\prec})^{\dag}=\tau\circ R_{q}^{\prec}=(R_{q}^{\succ})^{-1}\quad{\rm for}\;\;U_{q}(\mathfrak{su}(1,1)),&
\\[5pt]
&(R_{q}^{\succ})^{\dag}\;=\;(R_{q}^{\succ})^{{-1}},\quad (R_{q}^{\prec})^{\dag}\;=\;(R_{q}^{\prec})^{-1}\quad{\rm for}\;\;U_{q}(\mathfrak{sl}(2;\mathbb{R})),&
\end{array}
\end{eqnarray}
i.e. in the case $U_{q}(\mathfrak{sl}(2;\mathbb{R}))$ both $R$-matrices $R_{q}^{\succ}$, $R_{q}^{\prec}$ are unitary and in the case $U_{q}(\mathfrak{su}(1,1))$ they can be called ''flip-Hermitian'' or ''$\tau$-Hermitian''. 

In the limit $\alpha\rightarrow0$ ($q\rightarrow1$) we obtain for the $R$-matrix (\ref{qu5})
\begin{eqnarray}\label{qu13}
\begin{array}{rcl}
R_{q}(\mathfrak{g})\!\!&=\!\!&1+r_{BD}^{}+\textit{O}(\alpha^{2}).
\end{array}
\end{eqnarray}
Here $r_{BD}^{}$ is the classical Belavin-Drinfeld $r$-matrix:
\begin{eqnarray}\label{qu14}
\begin{array}{rcl}
r_{BD}^{}\!\!&=\!\!&2\beta\bigl(X_{+}^{}\otimes X_{-}^{}+X_{0}^{}\otimes X_{0}^{}\bigr),
\end{array}
\end{eqnarray}
where $\beta=\alpha$, $X_{\pm}^{}=E_{\pm}^{}$, $X_{0}^{}=H$ for the case $\mathfrak{g}=\mathfrak{su}(1,1)$, and $\beta=\imath\alpha$, $X_{\pm}=E_{\pm}'$, $X_{0}=H'$ for the case $\mathfrak{g}=\mathfrak{sl}(2;\mathbb{R})$. The $r$-matrix $r_{BD}^{}$ is not skew-symmetric and it satisfies the standard CYBE 
\begin{eqnarray}\label{qu15}
[r_{BD}^{12},r_{BD}^{13}+r_{BD}^{23}]+[r_{BD}^{13},r_{BD}^{23}]=0 
\end{eqnarray}
which is obtained from QYBE (\ref{qu6}) in the limit (\ref{qu13}). The standard $r$-matrix (\ref{is11}) or (\ref{is12}) is  the skew-symmetric part of $r_{BD}^{}$, namely
\begin{eqnarray}\label{qu16}
r_{BD}^{}\!\!&=\!\!&\frac{1}{2}\bar{r}_{st}^{}+\frac{1}{2}\bar{C}_{2}
\end{eqnarray}
where $\bar{r}_{st}^{}=r_{BD}^{12}-r_{BD}^{21}$ is the standard $r$-matrix (\ref{is11}) or (\ref{is12}) and $\bar{C}_{2}=2\beta C_{2}=r_{BD}^{12}+r_{BD}^{21}$ where $C_{2}$ is the split Casimir element of $\mathfrak{su}(1,1)$ or $\mathfrak{sl}(2;\mathbb{R})$.

We can introduce the quantum Cartesian generators by the formulas: $X_{\pm}^{}=\imath J_{1}^{}\pm J_{3}^{},\;q^{\pm X_{0}}=q^{\pm\imath J_{2}}.$\footnote{The generators $J_{i}^{}=(-1)^{i-1}J_{i}^{\dag}$ $(i=1,2,3)$ are $q$-analoqs of the Cartesian basis (\ref{pr3}), (\ref{pr5}) ($\lim_{q\to1} J_{i}^{}\to I_{i}^{}$).} In terms of these generators the quantum algebra $U_{q}(\mathfrak{su}(1,1))$, which will be denoted by $U_{(r_{st}^{})}(\mathfrak{o}(2,1))$, can be reformulated as follows. The quantum deformation of  $U(\mathfrak{o}(2,1))$, corresponding to the classical $r$-matrix (\ref{is11}), is an unital associative algebra $U_{(r_{st}^{})}(\mathfrak{o}(2,1))$ with the generators  $\{J_{1}^{},\;J_{3}^{},\;q^{\pm\imath J_{2}}\}$ and the defining relations $(k=1,3)$:
\begin{eqnarray}\label{qu17}
\begin{array}{rcl}
q^{\imath J_{2}}q^{-\imath J_{2}}\!\!\!&=\!\!&q^{-\imath J_{2}}q^{\imath J_{2}}=1,\quad[J_{1}^{},\,J_{3}^{}]\;=\;\displaystyle\frac{\imath(q^{2\imath J_{2}^{}}-q^{-2\imath J_{2}^{}})}{2(q-q^{-1})},
\\[8pt]
q^{\pm\imath J_{2}}J_{k}^{}\!\!\!&=\!\!&\displaystyle\frac{1}{2}(q+q^{-1})J_{k}^{}q^{\pm\imath J_{2}}\pm\frac{\imath}{2}(q-q^{-1})\varepsilon_{2kl}J_{l}^{}q^{\pm\imath J_{2}}
\end{array}
\end{eqnarray}
with the reality condition $J_{1}^{\dagger}=J_{1}^{}$, $J_{3}^{\dagger}=J_{3}^{}$, $(q^{\pm\imath J_{2}})^{\dag}=q^{\pm\imath J_{2}}$, $q^{*}=q$ ($q:=e^{\alpha}$, $\alpha\in\mathbb{R}$). These relations are the $q$-analog of the relations (\ref{pr3}) with the reality condition (\ref{pr5}). The Hopf algebra structure on $U_{(r_{st}^{})}(\mathfrak{o}(2,1))$ is given as follows $(k=1,3)$:
\begin{eqnarray}\label{qu18}
\begin{array}{rcl}
\Delta_{q}(J_{k}^{})\!\!&=\!\!&J_{k}^{}\otimes q^{\imath J_{2}}+q^{-\imath J_{2}}\otimes J_{k}^{},
\\[8pt]
\Delta_{q}(q^{\pm\imath J_{2}})\!\!\!&=\!\!&q^{\pm\imath J_{2}}\otimes q^{\pm\imath J_{2}},\quad S_{q}(q^{\pm\imath J_{2}})\;=\;q^{\mp\imath J_{2}},
\\[5pt] 
S_{q}(J_{k}^{})\!\!\!&=\!\!&-\displaystyle\frac{1}{2}(q+q^{-1})J_{k}^{}+\frac{\imath}{2}(q-q^{-1})\varepsilon_{k2l}J_{l}^{},
\\[9pt]
\epsilon_{q}(q^{\pm\imath J_{2}})\!\!&=\!\!&1,\quad\epsilon_{q}(J_{k}^{})\;=\;0, 
\end{array}
\end{eqnarray}
Substituting in the formulas (\ref{qu7}) and (\ref{qu9}) the expressions $X_{\pm}^{}=\imath J_{1}^{}\pm J_{3}^{},\;q^{\pm X_{0}}=q^{\pm\imath J_{2}}$ we obtain the universal $R$-matrix in the terms of the quantum Cartesian generators $J_{i}$ ($i=1,2,3$) with the defining relations (\ref{qu17}).
 
We can also introduce another quantum Cartesian generators by the formulas: $X_{\pm}^{}=\imath J_{1}'\mp J_{2}',\;q^{\pm X_0}=q^{\pm\imath J_{3}'}$.\footnote{The generators $J_{i}'$ $(i=1,2,3)$ are also the $q$-analoq of the Cartesian basis given by (\ref{pr3}), (\ref{pr5}) ($\lim_{q\to1} J_{i}'\to I_{i}^{}$).} In terms of these generators the quantum algebra $U_{q}(\mathfrak{sl}(2;\mathbb{R})$, which will be denoted by $U_{(r_{st}')}(\mathfrak{o}(2,1))$, can be reformulated as follows. The quantum deformation of  $U(\mathfrak{o}(2,1))$, corresponding to the classical $r$-matrix (\ref{is12}), is an unital associative algebra $U_{(r_{st}')}(\mathfrak{o}(2,1))$ with the generators  $\{J_{1}',\;J_{2}',\;q^{\pm\imath J_{3}'}\}$ and the defining relations $(k=1,2)$:
\begin{eqnarray}\label{qu19}
\begin{array}{rcl}
q^{\imath J_{3}'}q^{-\imath J_{3}'}\!\!\!&=\!\!&q^{-\imath J_{3}'}q^{\imath J_{3}'}=1,\quad[J_{1}',\,J_{2}']\;=\;\displaystyle-\frac{\imath(q^{2\imath J_{3}'}-q^{-2\imath J_{3}'})}{2(q-q^{-1})},
\\[10pt]
q^{\pm\imath J_{3}'}J_{k}'\!\!\!&=\!\!&\displaystyle\frac{1}{2}(q+q^{-1})J_{k}'q^{\pm\imath J_{3}'}\pm\frac{\imath}{2}(q-q^{-1})\varepsilon_{3kl}J_{l}'q^{\pm\imath J'_{3}}
\end{array}
\end{eqnarray}
with the reality conditions ${J_{1}'}^{\dagger}=J_{1}'$, ${J_{2}'}^{\dagger}=-J_{2}'$, $(q^{\imath J_{3}'})^{\dagger}=q^{\imath J_{3}'}$, $q^{*}=q^{-1}$ ($q:=e^{\imath\alpha}$, $\alpha\in\mathbb{R}$). The Hopf structure on $U_{(r_{st}')}(\mathfrak{o}(2,1))$ are provided by the formulae $(k=1,2)$:
\begin{eqnarray}\label{qu20}
\begin{array}{rcl}
\Delta_{q}(J_{k}')\!\!&=\!\!&J_{k}'\otimes q^{\imath J_{3}'}+q^{-\imath J_{3}'}\otimes J_{k}',
\\[8pt]
\Delta_{q}(q^{\pm\imath J_{3}'})\!\!\!&=\!\!&q^{\pm\imath J_{3}'}\otimes q^{\pm\imath J_{3}'},\quad S_{q}(q^{\pm\imath J_{3}'})\;=\;q^{\mp\imath J_{3}'},
\\[5pt] 
S_{q}(J_{k}')\!\!\!&=\!\!&-\displaystyle\frac{1}{2}(q+q^{-1})J_{k}'+\frac{\imath}{2}(q-q^{-1})\varepsilon_{k3l}J_{l}',
\\[9pt]
\epsilon_{q}(q^{\pm\imath J_{3}'})\!\!&=\!\!&1,\quad\epsilon_{q}(J_{k}')\;=\;0, 
\end{array}
\end{eqnarray}
Substituting in the formulas (\ref{qu7})) and (\ref{qu9})) the expressions $X_{\pm}^{}=\imath J_{1}'\pm J_{2}',\;q^{\pm X_{0}}=q^{\pm\imath J_{3}'}$ we obtain the universal $R$-matrix in terms of the quantum physical generators $J_{i}'$ ($i=1,2,3$) with the defining relations (\ref{qu19}).

The quantization of $U(\mathfrak{sl}(2;\mathbb{R}))$ corresponding to the classical Jordanian $r$-matrix (\ref{is13}) is well known for a long time \cite{GeGiSch1990,Og1993,KuLyMu1998} and it is defined by the twist $F$ (see \cite{Og1993}):
\begin{eqnarray}\label{qu21}
F\!\!&=\!\!&\exp(H'\otimes\sigma),\quad\sigma\,=\,\ln(1+\imath\alpha E_{+}').
\end{eqnarray}
The two-tensor $F$ satisfies the 2-cocycle condition
\begin{equation}\label{qu22}
F^{12}(\Delta\otimes{\rm id})(F)\;=\;F^{23}({\rm id}\otimes\Delta)(F),
\end{equation}
and the "unital" normalization
\begin{equation}\label{qu23}
(\epsilon\otimes{\rm id})(F)\;=\;({\rm id}\otimes\epsilon )(F)\;=\;1.
\end{equation}
It is evident that the twist (\ref{qu21}) is unitary
\begin{eqnarray}\label{qu24}
F^{*}\!\!&=\!\!&F^{-1}.
\end{eqnarray}
The twisting element $F$ defines a deformation of the universal enveloping algebra $U(\mathfrak{sl}(2;\mathbb{R}))$ considered as a Hopf algebra. The new deformed coproduct and antipode are given as follows
\begin{equation}\label{qu25}
\Delta^{(F)}(X)\;=\;F\Delta(X)F^{-1}~,\qquad S^{(F)}(X)=uS(X)u^{-1}
\end{equation}
for any $X\in U(\mathfrak{sl}(2;\mathbb{R}))$, where $\Delta(X)$ and $S(X)$ are the coproduct and the antipode before twisting: $\Delta(X)=X\otimes1+1\otimes X$, $S(X)=-X$; and
\begin{eqnarray}\label{qu26}
u\!\!&=\!\!&m({\rm id}\otimes S)(F)=\exp(-\imath\alpha H'E_{+}').
\end{eqnarray}
It is easy to see that we get the $*$-Hopf algebra, i.e.
\begin{eqnarray}\label{qu27}
(\Delta^{(F)}(X))^{*}\!\!&=\!\!&\Delta^{(F)}(X^{*}),\quad(S^{(F)}(X))^{*}=S^{(F)}(X^{*})
\end{eqnarray}
for any $X\in U(\mathfrak{sl}(2;\mathbb{R}))$. One can calculate the following formulae for the deformed coproducts $\Delta^{(F)}$ (see \cite{Og1993}):
\begin{eqnarray}\label{qu28}
\begin{array}{rcl}
\Delta^{(F)}(H')\!\!&=\!\!&H^{'}\otimes e^{-\sigma}+1\otimes H',
\\[5pt] 
\Delta^{(F)}(E_{+}')\!\!&=\!\!&E_{+}'\otimes e^{\sigma}+1\otimes E_{+}',
\\[5pt]
\Delta^{(F)}(E_{-}')\!\!&=\!\!&E_{-}'\otimes e^{-\sigma}+1\otimes E_{-}'+2\imath\alpha H'\otimes H'e^{-\sigma}
\\[4pt]
&&+\alpha^{2}H'(H'-1)\otimes E_{+}'e^{-2\sigma}.
\end{array}
\end{eqnarray}
Using (\ref{qu25}) and (\ref{qu26}) one gets the formulae for the deformed antipode $S^{(F)}$:
\begin{eqnarray}\label{qu29}
\begin{array}{rcl}
S^{(F)}(H')\!\!&=\!\!&-H^{'}e^{-\sigma},\quad S^{(F)}(E_{+}')\,=\,-E_{+}'e^{-\sigma},
\\[5pt] 
S^{(F)}(E_{-}')\!\!&=\!\!&-E_{-}'e^{\sigma}+2\imath\alpha{H'}^{2}e^{\sigma}
-\alpha^{2}H'(H'-1)E_{+}'e^{\sigma}.
\end{array}
\end{eqnarray}
It is easy to see the universal $R$-matrix $R^{(F)}$ for this twisted deformation looks as follows
\begin{eqnarray}\label{qu30}
R^{(F)}\!\!&=\!\!&\tilde{F}F^{-1},\quad (R^{(F)})^{*}=(R^{(F)})^{-1}.
\end{eqnarray}
In the limit $\alpha\rightarrow0$  we obtain for the $R$-matrix (\ref{qu23})
\begin{eqnarray}\label{qu31}
\begin{array}{rcl}
R^{(F)}\!\!&=\!\!&1+r_{J}^{}+\textit{O}(\alpha^{2}),
\end{array}
\end{eqnarray}
where $r_{J}^{}$ is the classical Jordanian $r$-matrix (\ref{is13}). Using the relations (\ref{pr6}) we can express all the formulas (\ref{qu28})--(\ref{qu30}) in terms of the Cartesian basis (\ref{pr3})) and (\ref{pr5})). 

We add that the Jordanian deformation has been described as well in a deformed $\mathfrak{sl}(2;\mathbb{R})$ algebra basis \cite{Ohn1992,Vl1993}.

\section{Short Summary and Outlook}
By using the three-fold isomorphism of classical Lie algebras $\mathfrak{o}(2,1) \simeq\mathfrak{sl}(2;\mathbb{R})\simeq\mathfrak{su}(1,1)$ one can express the infinitesimal versions of the $D=3$ Lorentz quantum deformations in terms of   classical  $\mathfrak{o}(2,1)$, $\mathfrak{sl}(2;\mathbb{R})$ and $\mathfrak{su}(1,1)$ $r$-matrices. The first aim of our paper was to derive $\mathfrak{o}(2,1)$, $\mathfrak{su}(1,1)$ and $\mathfrak{sl}(2;\mathbb{R})$ bialgebras using representation-independent purely algebraic method (see Sect.~3) and further	to provide the explicite maps which relate them (see Sect.~4). We start in Sect.~3  with the derivation of known pair of inequivalent complex $\mathfrak{o}(3;\mathbb{C})\simeq\mathfrak{sl}(2;\mathbb{C})$ $r$-matrices - the Jordanian (nonstandard) one and the Drinfeld-Jimbo (standard) $r$-matrix. Passing from $\mathfrak{sl}(2;\mathbb{C})$ to $\mathfrak{sl}(2;\mathbb{R})$ we obtain three independent $\mathfrak{sl}(2;\mathbb{R})$ $r$-matrices. First two of them are  the real forms of two  basic complex $\mathfrak{sl}(2;\mathbb{C})$ $r$-matrices, the third $\mathfrak{sl}(2;\mathbb{R})$ $r$-matrix, which we called quasi-standard (see \ref{rm46})), is the sum of two skew-symmetric 2-tensors. We do not know however how to obtain directly the universal $R$-matrix from the third $r$-matrix. We show that there is however a way out (see Sect.~4): the quasi-standard $r$-matrix (\ref{rm46})) (see also (\ref{rm7})) can be transformed by the map (\ref{pr11}) into the standard $r$-matrix in $\mathfrak{su}(1,1)$ basis, with known universal $R$-matrix (see e.g. \cite{Ma1995}). In such a way we can derive the effective quantization of all three $D=3$ Lorentz $r$-matrices, however we recall that for such a purpose it is necessary to use both $\mathfrak{sl}(2;\mathbb{R})$ and $\mathfrak{su}(1,1)$ bases. 
	
In second part of Introduction we mentioned main applications of $D=3$ Lorentz symmetries and their deformations, but still more important for the description of noncommutative $D=3 $ space-time geometry and $D=3$ quantum gravity are the quantum deformations of $D=3$ Poincar\'{e} algebra, with noncommutative translations sector. These quantum deformations were classified  (see e.g. \cite{St1998}) in terms of classical $r$-matrices, but systematic studies of their Hopf quantizations still should be completed. There were considered also the quantum deformations of $D=3$ de-Sitter ($dS$) and anti-de-Sitter ($AdS$) space-times, with nonvanishing cosmological constant $\Lambda$. In $D=3$ $dS$ case ($\Lambda>0$) all Hopf-algebraic quantizations are known, because they were studied as the quantum deformations of $D=4$ Lorentz algebra $\mathfrak{o}(3,1)$ \cite{BoLuTo2008}. In $D=3$ $AdS$ case ($\Lambda<0$) with $\mathfrak{o}(2,2)$ symmetry some Hopf-algebraic quantum deformations were also given, but  recently there was presented complete classification of real $\mathfrak{o}(2,2)$ $r$-matrices\footnote{See \cite{BoLuTo2016} and Addendum (to be published); for earlier efforts to describe $\mathfrak{o}(2,2)$ quantum deformations see e.g.  \cite{BaHeOlSa1995}. We recall that $\mathfrak{o}(D,D)$ algebras describe  the symmetries of double geometry \cite{BeTh2008, HoLuZw2013}, which were used recently  e.g. in the description of self-dual models in so-called metastring theory \cite{FeLeMi2005}.}.
	
For physical applications it is very important  to consider subsequently the  quantum space-time deformations for $D=4,5,6$. The deformations of physical $D=4$ space-time and $D=4$ Poincar\'{e} algebra were extensively studied for more than a quarter of the century \cite{CaSchlSchoWa1991}--\cite{Zak1997}, but it should be observed that the complete list of $D=4$ Poincar\'{e} $r$-matrices ($D=4$ Poincar\'{e} bialgebras) is still not complete\footnote{The classification problems of $D=4$ Poincar\'{e} bialgebras described in \cite{Zak1997} still remain unsolved.}. The next task could be to describe all deformations of $D=4$ space-times with constant curvature and arbitrary signature, which would classify all possible $D=4$ quantum $dS$ and $AdS$ algebras as well as the quantum-deformed $D=5$ Euclidean $\mathfrak{o}(5)$ symmetries. For such a purpose one can look for the extension of algebraic methods used to classify the deformations of $\mathfrak{o}(4;\mathbb{C})$ and its real forms  (see \cite{BoLuTo2016}) to the case of $\mathfrak{o}(5;\mathbb{C})$ and the real forms $\mathfrak{o}(5)$, $\mathfrak{o}(4,1)$ and $\mathfrak{o}(3,2)$. Finally the systematic study of deformations of $\mathfrak{o}(6;\mathbb{C})$ is another important challege, in particular because the deformations of its real form $\mathfrak{o}(4,2)\simeq\mathfrak{su}(2,2)$ will provide the list of quantum $D=4$ conformal algebras. 

\subsection*{Acknowledgments}
The authors would like to thank A.~Borowiec for discussions. First author (JL) would like to acknowledge the financial support of NCN (Polish National Science Center) grants 2013/09/B/ST2/02205, 2014/13/B/ST2/04043 and by European Project COST, Action MP1405 QSPACE. Second author (VNT) is supported by RFBR grant No. 16-01-00562-a.



\begin{thebibliography}{99} 

\bibitem{FrLi2006} L. Freidel, E.R. Livine, \textit{Ponzano-Regge model revisited III: Feynman diagrams and Effective field theory,	Class. Quant. Grav.} \textbf{23} (2006) 2021, arXiv:hep-th/0502106v2; \textit{3d Quantum Gravity and Effective Non-Commutative Quantum Field Theory, Phys. Rev. Lett.} \textbf{96} (2006) 221301, arXiv:hep-th/0512113v2.

\bibitem{CiKoPrRo2016} F.~Cianfrani, J.~Kowalski-Glikman, D.~Pranzetti, G.~Rosati, \textit{Symmetries of quantum space-time in 3 dimensions, Phys. Rev.}  \textbf{D94} No. 8 (2016) 084044, arXiv:1606.03085 [hep-th].

\bibitem{Dr1986} V.~Drinfeld, \textit{Quantum Groups, Proc. Int. Congress of Math. Berkeley, Academic Press} \textbf{1} (1986) 798.

\bibitem{EtKa1996} P. Etingof and D. Kazhdan, \textit{Quantization of Lie bialgebras, I, Selecta Mathematica, New Series} \textbf{2} (1996) 1, arXiv:q-alg/9506005v5.  

\bibitem{ChPr1994}V.~Chari, A.~Pressley, \textit{``A Guide to Quantum Groups'', Cambridge Univ. Press} 1994.

\bibitem{Ma1995} S.~Majid, \textit{``Foundations of Quantum Groups'', Cambridge Univ. Press} (1995).

\bibitem{BoLuTo2016} A.~Borowiec, J.~Lukierski, V.N.~Tolstoy, \textit{Quantum deformations of $D=4$ Euclidean, Lorentz, Kleinian and quaternionic $\mathfrak{o}^*(4)$ symmetries in unified $\mathfrak{o}(4;C)$ setting, Phys. Lett.} \textbf{B754} (2016) 176, arXiv:1511.03653 [hep-th].

\bibitem{Re1996} A.G. Reyman, \textit{Poisson structures related to quantum groups, in ''Quantum Groups and its Applications in Physics'', Jnt. School ''Enrico Fermi'', Varrena 1994}, eds. L. Castellani and J. Wess, IOS, Amsterdam (1996), p.407.  

\bibitem{Go2000} X.~Gomez, \textit{Classification of three-dimensional Lie bialgebras, J. Math. Phys.} \textbf{41} (2000) 4939.

\bibitem{ReHeRa2005} A.~Rezaei-Aghdam, M.~Hemmati, A.R.~Rastkar, \textit{Classification of real three-dimensional Lie bialgebras and their Poisson-Lie groups, J. Phys. A: Math. Gen.} \textbf{A38} (2005) 3981, arXiv:math-ph/0412092. 

\bibitem{BaBlMu2012} A.~Ballesteros, A.~	Blasco, F.~Musso, \textit{Classification of real three-dimensional Poisson-Lie groups, J. Phys. A: Math. Theor.} \textbf{A45} (2012) 175204, arXiv:1202.2077 [math-ph]. 

\bibitem{BaMeNa2017} A.~Ballesteros, C.~Meusburger, P.~Naranjo, \textit{$AdS$ Poisson homogeneous spaces and Drinfeld doubles}, arXiv:1701.04902 [math-ph]. 

\bibitem{BoHaReSe2011} R.~Borcherds, M.~Haiman, N.~Reshetikhin, V.~Serganova, \textit{``Berkeley Lecture on Lie Groups and Quantum Groups'', ed. Anton Geraschenko and Theo Johnson-Freyd.} Last updated September 22, 2011; http://math.berkeley.edu/~theojf/LieQuantumGroups.pdf; see N.~Reshetikhin, part II \textit{Quantum groups,} Sect.~9.4.4.

\bibitem{AlFuFu1976} V.~de Alfaro, S.~Fubini, P.~Furlan, \textit{Conformal invariance in Quantum Mechanics, Nuovo Cim.} \textbf{A34} (1976) 569.

\bibitem{IvKrLe1989} E.~Ivanov, S.O.~Krivonos, V.M.~Leviant, \textit{Geometry of Conformal Mechanics, J. Phys.} \textbf{A22} (1989) 345.

\bibitem{FeIvLu2011} E.~Fedoruk, E.~Ivanov, J.~Lukierski, \textit{Galilean Conformal Mechanics from Nonlinear Realizations, Phys. Rev.} \textbf{D83} (2011) 085013, arXiv:1101.1658 [hep-th].

\bibitem{BoLuTo2003} A.~Borowiec, J.~Lukierski, V.N.~Tolstoy, \textit{Basic Twist Quantization of $osp(1|2)$ and $\kappa$-Deformation of $D=1$ Superconformal Mechanics, Mod. Phys. Lett.} \textbf{A18} (2003) 1157,  arXiv:hep-th/0301033.

\bibitem{AchTo1986} A.~Achucarro, P.~Townsend, \textit{A Chern-Simons action for three-dimensional anti-de Sitter supergravity theories, Phys. Lett.} \textbf{B180} (1986) 89.

\bibitem{Wi1988} E.~Witten, \textit{(2+1)-dimensional gravity as an exactly soluble system, Nucl. Phys.} \textbf{B311}, 46(1988).

\bibitem{Ca1998} S.~Carlip, \textit{''Quantum Gravity in $2+1$ Dimention'', Cambridge University Press}, 1998.

\bibitem{MeSch2003} C.~Meusburger, B.J.~Schroers, \textit{Poisson structure and symmetry in the Chern-Simons formulation of (2+1)-dimensional gravity, Class. Quant. Grav.} \textbf{20} (2003)  2193, \hbox{gr-qc/0301108}. 

\bibitem{MeSch2009} C.~Meusburger, B.J.~Schroers, \textit{Generalized Chern-Simons actions for $3d$ gravity and $\kappa$-Poincare symmetry, Nucl. Phys.} \textbf{B806} (2009) 462, arXiv:0805.3318 [gr-qc].

\bibitem{BaHeMu2014} A.~Ballesteros, F.J.~Herranz, F.~Musso, \textit{On quantum deformations of (anti-)de Sitter algebras in (2+1) dimensions, J. Phys. Conf. Serie} \textbf{532} (2014) 012002, arXiv:1302.0684 [hep-th].

\bibitem{PoRe1968} G.~Ponzano, T.~Regge, \textit{in: Spectroscopic and group-theoretical methods in physics} (1968) 1, ed. F.~Bloch, North-Holland Publ. Co.

\bibitem{Li2001} E.R.~Livine, \textit{Short and Subjective Introduction to the Spinfoam Framework for Quantum Gravity}, arXiv:1101.5061 [gr-qc].

\bibitem{ArJu2001} I~Ambj\"{o}rn, J.~Jurkiewicz, R.~Loll, \textit{Dynamically Triangulating Lorentzian Quantum Gravity, Nucl. Phys.} \textbf{B610} (2001) 347, arXiv:hep-th/0105267.

\bibitem{MaMiNaNoUe1990} T.~Masuda, K.~Mimachi, Y.~Nakagami, M.~Noumi, Y.~Saburi, K.~Ueno, \textit{Unitary representations of the quantum group $SU_{q}(1,1)$: Structure of the dual space of $U_{q}(sl(2))$, Lett. Math. Phys.} \textbf{19} (1990) 187, \textit{Unitary representations of  the quantum group $SU_{q}(1,1)$: Matrix elements of unitary representations and the basic hypergeometric functions, Lett. Math. Phys.} \textbf{19} (1990) 195.

\bibitem{Lax1968} P.~Lax, \textit{Commuting Integrals of nonlinear equations of evolution and solitary waves, Pure Appl. Math.} \textbf{21} (1968) 467.

\bibitem{Bl1998} M.~B\l aszak, \textit{''Multi-Hamiltonian Theory of Dynamical Systems''}, Springer (1998).

\bibitem{MeTs1998} R.R.~Metsaev, A.A.~Tseytlin, \textit{Type IIB superstring action in $AdS_5\times S^5$ background, Nucl. Phys.} \textbf{B533} (1998) 109, arXiv:hep-th/9805028. 

\bibitem{Kl2002} C.~Klim\v{c}\'{\i}k, \textit{Yang-Baxter $\sigma$-models and $dS/AdS$ $T$-duality, JHEP} \textbf{12} (2002) 051, 28. arXiv:hep-th/0210095; \textit{On integrability of the Yang-Baxter $\sigma$-model, J. Math. Phys.} \textbf{50} (2009) 043508, arXiv:0802.3518 [hep-th]. 

\bibitem{DeMaVi2016} F.~Delduc, M.~Magro, B.~Vicedo, \textit{An integrable deformation of the $AdS_5\times S^5$ superstring action, Phys. Rev. Lett.} \textbf{112} (2014) 051601, arXiv:1309.5850 [hep-th].

\bibitem{KaMaYo2014} I.~Kawaguchi, T.~Matsumoto, K.~Yoshida, \textit{Jordanian deformations of the $AdS_5\times S^5$ superstring, JHEP} \textbf{04} (2014) 153, arXiv:1401.4955 [hep-th].

\bibitem{ArFrHoRoTs2016} G. Arutyunov, S. Frolov, B. Hoare, R. Roiban, A.A. Tseytlin, \textit{Scale invariance of the $\eta$-deformed $AdS5\times S5$ superstring, $T$-duality and modified type II equations, Nucl. Phys.} \textbf{B903} (2016) 262, arXiv:1511.05795 [hep-th].

\bibitem{Ho2015} B.~Hoare, \textit{Towards a two-parameter $q$-deformation of $AdS_3\times S^3\times M^4$ superstrings, Nucl. Phys.} \textbf{B891} (2015) 259, arXiv:1411.1266 [hep-th].

\bibitem{OsTo2017} D. Osten, S. van Tongeren, \textit{Abelian Yang-Baxter Deformations and $TsT$ transformations, Nucl.Phys.} \textbf{B915} (2017) 184, arXiv:1608.08504 [hep-th].

\bibitem{ChLu2006} Y.~Chervonyi, O.~Lunin, \textit{Supergravity background of the lambda-deformed $AdS_3\times S^3$ supercoset} arXiv:1606.00934 [hep-th].

\bibitem{KlSch1997} A.~Klimyk, K.~Schmudgen, \textit{Quantum Groups and their Representations, Springer-Verlag, Berlin Heidelberg} (1997) p.~58.

\bibitem{KhTo1991} S.M.~Khoroshkin and V.N.~Tolstoy, \textit{Universal $R$-matrix for quantized (super)algebra, Comm. Math. Phys.} \textbf{141} (1991) 599.

\bibitem{KhTo1992} S.M.~Khoroshkin and V.N.~Tolstoy, \textit{The uniqueness theorem for the universal $R$-Matrix, Lett. in Math. Phys.} \textbf{24} (1992) 231.

\bibitem{GeGiSch1990} M.~Gerstenhaber, A.~Giaquinto, S.D.~Schack, \textit{Quantum symmetry, in: ``Quantum Groups'', Proc. of EIMI Workshop 1990, Leningrad, ed. P.P.Kulish, Springer Lecture Notes in Math.} \textbf{1510} (1992) 9.

\bibitem{Og1993} O.V.~Ogievetsky, \textit{Hopf structures on the Borel subalgebra of $sl(2)$, in Proc. Winter School ``Geometry and Physics'',  Zidkov, January 2013, Czech Republic, Rendiconti Circ. Math. Palermo, Serie II} \textbf{37} (1993) 185, Max Planck Int. prepr. MPI-Ph/92-99.

\bibitem{KuLyMu1998} P.P.~Kulish, V.D.~Lyakhovsky, A.I.~Mudrov, \textit{Extended Jordania twists for Lie algebras, J. Math. Phys.} \textbf{40} (1999) 4569, math/9806014.

\bibitem{Ohn1992} C.~Ohn, \textit{A *-product on $SL(2)$ and the corresponding nonstandard (sl(2))” Lett. Math. Phys.} \textbf{25} (1992) 85.

\bibitem{Vl1993} A.A.~Vladimirov, \textit{On the Hopf algebras generated by the Yang-Baxter R-matrices, Z. Phys.} \textbf{C58} (1993) 659, hep-th/9302043; \textit{A closed expression for the universal $R$-matrix in a nonstandard quantum double, Mod. Phys. Lett.} \textbf{A8} (1993) 2573, hep-th/9305048.

\bibitem{St1998} P.~Stachura, \textit{Poisson-Lie structures on Poincare and Euclidean groups in three dimensions, J. Phys.} \textbf{A31} (1998) 4555.

\bibitem{BoLuTo2008} A.~Borowiec, J.~Lukierski, V.N.~Tolstoy, \textit{Quantum deformations of $D=4$ Lorentz algebra revisited: twistings of $q$-deformation, Eur. Phys. J.} \textbf{C57} (2008) 601, arXiv:0804.3305 [hep-th].

\bibitem{BaHeOlSa1995} A.~Ballesteros, F.J.~Herranz, M.A.~del~Olmo, M.~Santander, \textit{Non-standard quantum $so(2,2)$ and beyond, J. Phys.} \textbf{A28} (1995) 941, hep-th/9406098. 

\bibitem{BeTh2008} D.S.~Berman, D.C.~Thompson, \textit{Duality Symmetric Strings, Dilatons and $O(d,d)$ Effective Actions, Phys. Lett.} \textbf{B662} (2008) 279, arXiv:0712.1121 [hep-th]; \textit{Duality Symmetric String and M-Theory, Phys. Rept.} \textbf{566} (2014) 1, arXiv:1306.2603 [hep-th].

\bibitem{HoLuZw2013} O.~Hohm, D.~ L\"ust, B.~Zwiebach, \textit{The Spacetime of Double Field Theory: Review, Remarks, and Outlook, Fortschr. Phys.} \textbf{61} (2013) 926, arXiv:1309.2977 [hep-th]. 

\bibitem{FeLeMi2005} L.~Freidel, R.G.~Leigh, D.~Minic, \textit{Quantum gravity, dynamical phase-space and string theory, Int. J. Mod. Phys.} \textbf{D23} (2014) 1442006, arXiv: 1405.3949; \textit{Metastring Theory and Modular Space-time}, arXiv:1502.08005.

\bibitem{CaSchlSchoWa1991} V.~Carow-Watamura, S.~ Schliecher, M. Scholl, S.~Watamura, \textit{Quantum Lorentz group, Jnt. J. Mod. Phys.} \textbf{6} 3081 (1991).

\bibitem{OgScWeZu1992} O.V.~Ogievetsky, W.B. Schmidtke, J. Wess, B. Zumino, \textit{$q$-deformed Poincare algebra, Commun. Math. Phys.} \textbf{150} (1992) 495. 

\bibitem{LuRuNoTo1991} J.~Lukierski, A.~Nowicki, H.~Ruegg, V. N.~Tolstoy, \textit{$q$-deformation of Poincar ́e algebra, Phys. Lett.} \textbf{B264} (1991) 331. 

\bibitem{Ma1993} S.~Majid, \textit{Braided momentum in the q-Poincare group, J. Math. Phys.} \textbf{34} (1993) 2045.

\bibitem{Zak1997} S.~Zakrzewski, \textit{Poisson structures on the Poincare group, Commun. Math. Phys.} \textbf{185} (1997) 285, arXiv:q-alg/9602001.

\end{thebibliography}
\end{document}